\documentclass[aps,prl,twocolumn,amsmath,amssymb,superscriptaddress]{revtex4-2}
\usepackage{graphicx} 
\usepackage{xcolor}
\usepackage{threeparttable}
\usepackage[colorlinks=true,
					linkcolor=blue,
					anchorcolor=blue,
					citecolor=blue,
					urlcolor=magenta]{hyperref}
\usepackage{hyperref}
\usepackage[normalem]{ulem}

\newcommand{\ket}[1]{\ensuremath{\left|#1\right\rangle}}

\begin{document}


\title{Suppressing spurious transitions using spectrally balanced pulse}

\newcommand{\BAQIS}{\affiliation{1}{Beijing Key Laboratory of Fault-Tolerant Quantum Computing, Beijing Academy of Quantum Information Sciences, Beijing 100193, China}}

\newcommand{\IOP}{\affiliation{2}{Beijing National Laboratory for Condensed Matter Physics and Institute of Physics, Chinese Academy of Sciences, Beijing 100190, China}}

\newcommand{\UCAS}{\affiliation{3}{University of Chinese Academy of Sciences, Beijing 101408, China}}

\newcommand{\JULICH}{\affiliation{4}{Forschungszentrum J\"{u}lich, Institute of Quantum Control (PGI-8), D-52425 J\"{u}lich, Germany}}

\newcommand{\COLOGNE}{\affiliation{5}{Institute for Theoretical Physics, University of Cologne, D-50937 Cologne, Germany}}

\author{Ruixia Wang}
\email{wangrx@baqis.ac.cn}
\affiliation{\BAQIS}

\author{Yaqing Feng}
\affiliation{\BAQIS}

\author{Yujia Zhang}
\affiliation{\BAQIS}
\affiliation{\IOP}
\affiliation{\UCAS}
\author{Jiayu Ding}
\affiliation{\BAQIS}

\author{Boxi Li}
\affiliation{\JULICH}
\affiliation{\COLOGNE}
\author{Felix Motzoi}
\affiliation{\JULICH}
\affiliation{\COLOGNE}

\author{Yang Gao}
\affiliation{\BAQIS}
\affiliation{\IOP}
\affiliation{\UCAS}
\author{Huikai Xu}
\affiliation{\BAQIS}

\author{Zhen Yang}
\affiliation{\BAQIS}
\author{Wuerkaixi Nuerbolati}
\affiliation{\BAQIS}

\author{Haifeng Yu}
\affiliation{\BAQIS}
\author{Weijie Sun}
\email{sunwj@baqis.ac.cn}
\affiliation{\BAQIS}
\author{Fei Yan}
\email{yanfei@baqis.ac.cn}
\affiliation{\BAQIS}


\begin{abstract}

Achieving precise control over quantum systems presents a significant challenge, especially in many-body setups, where residual couplings and unintended transitions undermine the accuracy of quantum operations. In superconducting qubits, parasitic interactions---both between distant qubits and with spurious two-level systems---can severely limit the performance of quantum gates. In this work, we introduce a pulse-shaping technique that uses spectrally balanced microwave pulses to suppress undesired transitions. Experimental results demonstrate an order-of-magnitude reduction in spurious excitations between weakly detuned qubits, as well as a substantial decrease in single-qubit gate errors caused by a strongly coupled two-level defect over a broad frequency range. 
Our method provides a simple yet powerful solution for mitigating adverse effects from parasitic couplings, enhancing quantum gate fidelity. The pulse-shaping technique can be readily adapted to various physical systems.

\end{abstract}

\maketitle

\emph{Introduction} --- Quantum information processing technologies have made significant progress, achieving coherent control over systems with about 100 qubits \cite{Kim2023Evidence, acharya_quantum_2024, cao2023generation, xu2023digital, iqbal_non-abelian_2024, bluvstein2024logical}. In superconducting quantum processors, qubit-qubit interactions are typically mediated by engineered coupling elements, such as coplanar capacitors. However, due to the long-range nature of electromagnetic interactions and design constraints, parasitic or residual couplings can exist between qubits that are intended to be uncoupled. These parasitic interactions, known as quantum crosstalk \cite{Sarovar2020Detecting}, pose a challenge to the execution of independent operations, as they can interfere with transitions close in frequency. This phenomenon degrades the performance of quantum operations, limiting the scalability of quantum computing systems. 

Compounding this issue is the presence of uncontrolled microscopic degrees of freedom, often referred to as two-level defects or two-level systems (TLS), which are ubiquitous in many quantum platforms \cite{muller_towards_2019}. When frequencies of TLSs come close to those of the qubits, they can significantly interfere with qubit operations. Excitations into a long-lived TLS can be particularly harmful as they can accumulate over time and deteriorate subsequent operations, causing correlated or non-Markovian errors \cite{niu2019learning}. Without reset capability, this may be a potential threat to quantum error correction \cite{ghosh_understanding_2013}. Moving the qubit frequency away from these TLSs can reduce their impact. However, this approach is not directly available for fixed-frequency qubits. Even for tunable qubits, the available frequency options are often severely constrained by the presence of multiple TLSs, complicating the calibration of large-scale processors \cite{klimov_optimizing_2024}.

Various techniques have been developed to address these challenges \cite{bylander_noise_2011, ding2020systematic, murali2020software, zhao2023mitigation}. 
The Derivative Removal by Adiabatic Gate (DRAG) method \cite{Motzoi2009Simple, Gambetta2011Analytic, Chen2016Measuring, hyyppa2024reducing}, in particular, was a pulse shaping technique designed to suppress leakage transitions to higher energy states during single-qubit operations. By adding the derivative of the original pulse envelope to the quadrature component, DRAG removes unwanted diabatic leakage transitions, which, in the weak drive regime, show up as a spectral hole at a frequency determined by the prefactor of the derivative. 
The versatility of DRAG has been investigated in a variety of scenarios, including three-level lambda systems \cite{Vezvaee2023Avoiding}, cross-resonance gates \cite{Malekakhlagh2022Mitigating, Wei2024Characterizing, Li2023Suppression}, and frequency-crowded multilevel systems \cite{Theis2016Simultaneous}. 
However, traditional DRAG corrections face challenges when dealing with weakly detuned transitions, as they struggle to effectively remove spectral components close to the target transition \cite{Wei2024Characterizing}.

In this work, we propose and experimentally demonstrate a robust approach to mitigating undesired transitions caused by quantum crosstalk during single-qubit operations. Unlike conventional DRAG techniques, our method employs the dual-DRAG protocol, which creates symmetric spectral holes around the target transition frequency. This approach significantly reduces off-resonance effects during pulses and enables the pulse calibration to suppress unwanted transitions that are only slightly detuned. Residual off-resonance effects are further corrected using compensatory virtual-Z (VZ) gates. Experimental results validate the technique, demonstrating an order-of-magnitude suppression of crosstalk-induced excitations between two superconducting qubits detuned by approximately 40 MHz with a 25-ns pulse across various coupling strengths. Furthermore, we show its effectiveness in reducing excitations from a strongly coupled TLS and enhancing single-qubit gate fidelity across a broad frequency range, extending down to 20 MHz.

\begin{figure}[htbp]
\centering
\includegraphics[scale=1]{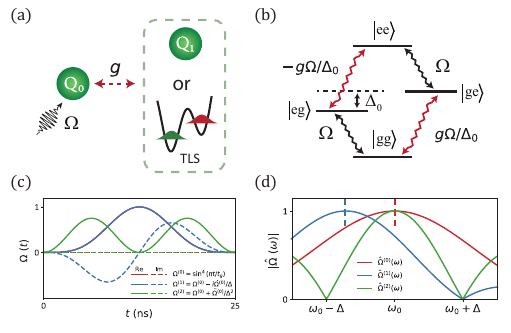}
\caption{\label{fig: scheme}
(a) Schematic diagram illustrating the unintended coupling ($g$) that can occur between two qubits or between a qubit and a spurious TLS.
(b) Energy level diagram for the combined system of $\rm Q_0$ and $\rm Q_1$ (or TLS). Due to state dressing, driving $\rm Q_0$ with an amplitude $\Omega$ also cross-drives $\rm Q_1$ with a reduced amplitude of $\pm g \Omega / \Delta_0$ in the weak coupling limit ($g \ll |\Delta_0|$). Cross-driving can induce unwanted transitions in $\rm Q_1$ when using a pulse with finite width.
(c) Time-domain pulse profiles and (d) their normalized Fourier spectra for a 25-ns pulse, showing its original sine-to-the-fourth-power form ($\rm \Omega^{(0)}$), the first-order DRAG-corrected pulse ($\rm \Omega^{(1)}$), and the spectrally balanced pulse with dual-DRAG corrections ($\rm \Omega^{(2)}$). Note that, in the time domain, the real part of $\Omega^{(0)}$ overlaps with that of $\Omega^{(1)}$, while the imaginary part of $\Omega^{(0)}$ overlaps with that of $\Omega^{(2)}$. The vertical dashed lines in (d) indicate the corresponding peak positions. In this case, the frequency to block is assumed to be $\rm \Delta/2\pi = 40\, MHz$. 
}
\end{figure}

\emph{Quantum crosstalk} --- Parasitic couplings are a common challenge in practical quantum hardware, often arising between qubits that are not intentionally connected. For example, in solid-state devices like superconducting quantum processors, qubits are typically designed for nearest-neighbor connections. However, the long-range nature of electromagnetic interactions can lead to unintended couplings between physically distant qubits, as shown in Fig.~\ref{fig: scheme}(a). These interactions can also occur through shared modes, such as chip or box modes. Additionally, unwanted couplings can exist between qubits and spurious quantum systems in the environment, such as TLSs. These uncontrollable systems present a major obstacle in the development and calibration of state-of-the-art quantum hardware.

Consider a parasitic exchange-type coupling $g$ between $\rm Q_0$ and $\rm Q_1$ (or a TLS). Due to the hybridization of the $\ket{\rm ge}$ and $\ket{\rm eg}$ states, driving one qubit inevitably induces a cross-driving effect on the other qubit or TLS, but with a reduced amplitude of $g\Omega/\Delta_0$, where $\Omega$ is the original drive amplitude applied to $\rm Q_0$, and $\Delta_0 = \omega_{\rm ge} - \omega_{\rm eg}$ is the detuning between the unwanted and target transitions, as illustrated in Fig.~\ref{fig: scheme}(b). Natural spectral broadening caused by finite pulse widths can lead to unwanted transitions, such as $\ket{\rm gg}\leftrightarrow\ket{\rm ge}$, which degrade the performance of quantum operations. This crosstalk phenomenon is inherently quantum mechanical; the crosstalk Hamiltonian is of the ZX type, exhibiting opposite signs depending on the state of $\rm Q_0$. This contrasts with classical signal crosstalk, where the Hamiltonian is of the IX type. As a result, quantum crosstalk cannot be simply corrected by applying a cancellation signal \cite{Nuerbolati2022Canceling}.
Interestingly, this crosstalk effect can be leveraged to construct a type of two-qubit entangling gate, the cross-resonance gate \cite{rigetti2010fully,de2010selective,Tripathi2019Operation}.

\emph{Dual-DRAG} --- A straightforward solution to suppress an unwanted transition is to remove the spectral components at that transition frequency. A famous example is the DRAG method, which adds a first-order derivative of the pulse envelope as a quadrature component, with an amplitude factor $1/\Delta$, where $\Delta$ is the DRAG parameter in angular frequency units. An example is illustrated in Fig.~\ref{fig: scheme}(c). This creates a spectral hole at frequency $\Delta$ relative to the drive frequency, as depicted in Fig.~\ref{fig: scheme}(d), effectively reducing spectral components near the qubit's $\ket{\rm e}$-$\ket{\rm f}$ transition — typically 200-300~MHz below the $\ket{\rm g}$-$\ket{\rm e}$ transition for transmon qubits — and suppressing leakage transitions.

Avoiding weakly detuned transitions during single-qubit operations, which are only a few tens of MHz away, has proven to be challenging \cite{Wei2024Characterizing}. 
For instance, as shown in Fig.~\ref{fig: scheme}(c,d) a 25-ns pulse with a single DRAG correction of $\Delta/2\pi = 40\, {\rm MHz}$ shifts the drive frequency by approximately 30 MHz.
When the DRAG-induced off-resonance effect is this large, the perturbative treatment in the DRAG theory breaks down, often causing the conventional mitigation approach which introduces a constant drive frequency detuning $\eta$ \cite{Chen2016Measuring} to fail.

To address this issue, we adopt the recursive DRAG scheme, where DRAG corrections are applied sequentially according to
\begin{equation} 
\Omega^{(n)} = \Omega^{(n-1)} - i \frac{\dot{\Omega}^{(n-1)}}{\Delta_n} \;, 
\label{eq:On} 
\end{equation} 
using a set of DRAG parameters $\{\Delta_1, \Delta_2, \dots, \Delta_n \}$ \cite{motzoi2013improving, Li2023Suppression}. Here, $\Omega^{(0)}$ represents the original pulse shape without DRAG. The final pulse $\Omega^{(n)}$ generates multiple spectral holes at the specified frequencies. By two successive DRAG applications at $\pm\Delta$, we obtain a pulse envelope that involves a second derivative,  
\begin{equation} 
\Omega^{(2)} = \Omega^{(0)} + \frac{\ddot{\Omega}^{(0)}}{\Delta^2} \;. 
\label{eq:DRAG2} 
\end{equation} 
As shown in Fig.~\ref{fig: scheme}(d), the spectrum $\hat{\Omega}^{(2)}(\omega)$ remains centered at $\omega_0$, while spectral components at $\omega_0\pm\Delta$ are eliminated. 
This dual-DRAG application at mirrored frequencies suppresses weakly detuned transitions and significantly reduces drive-induced off-resonance \cite{SM}. 
In the example presented, we choose the original pulse shape to be a sine raised to the fourth power. This ensures that the first three derivatives vanish at the pulse's beginning and end, thereby avoiding sharp edges or the need for truncation.

Notably, Ref.~\cite{hyyppa2024reducing} reports remarkable leakage suppression through their implementation of multiple derivative techniques. Their method, while not explicitly discussed, effectively utilizes an in-phase component analogous to Eq.~(\ref{eq:DRAG2}), applying successive DRAG corrections at $\pm\alpha$ followed by an additional correction at $\alpha$. Crucially, this arrangement naturally improves spectral symmetry---a fundamental property that, as we have identified, plays a pivotal role in mitigating DRAG-induced off-resonance effects and enabling their successful high-order derivative suppression.

\begin{figure}[t!]
\centering
\includegraphics[scale=1]{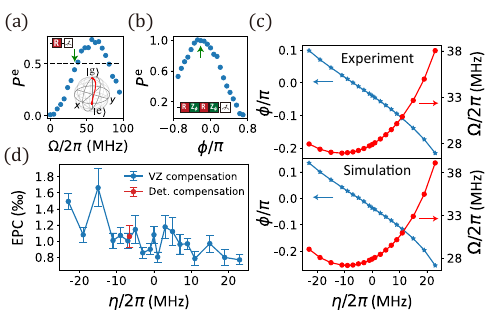}
\caption{\label{fig: vz}  
(a) Calibrating the pulse amplitude of the $\rm R$ gate ($t_g=25$\,ns) which drives the qubit from its ground state to an equator state, as indicated by the arrow. The drive detuning is set at $\eta/2\pi=23\,{\rm MHz}$. The inset plots are the circuit and a sketch of the state evolution on the Bloch sphere.
(b) Calibrating the compensating virtual-Z phase after the $\rm R$ gate, which yields an $\rm \sqrt{X}$ gate. By applying the combination twice, the qubit is rotated from the ground state to the excited state, as indicated by the arrow. The inset is the circuit.
(c) Experimental results (top panel) showing the calibrated drive amplitude and virtual-Z phase for different drive detunings, compared with numerical simulations (bottom panel).
(d) Measured error per Clifford (EPC) from randomized benchmarking using pulses with virtual-Z compensation (blue) for different drive detunings. Results using detuning compensation (without the appended virtual-Z phase, shown in red) are also included for comparison.}
\end{figure}

\emph{Correcting off-resonance error using virtual-Z compensation} --- Although the dual-DRAG method substantially reduces the deviation of spectral weighting, system nonidealities such as the presence of higher energy levels and pulse distortions can lead to residual off-resonance effects. To address this, we adopt a post-correction approach that compensates for the overall effect of an off-resonant pulse by appending a VZ gate. We validate this protocol on a device similar to that described in Ref.~\cite{chen2025efficient} (see Supplemental Material \cite{SM} for more information). For all experiments in this work unless specified otherwise, we constantly apply one DRAG correction at the qubit anharmonicity $\alpha/2\pi\approx-190\, {\rm MHz}$ to prohibit leakage transition to the $\ket{\rm f}$ state.

Throughout our work, we employ the U3 decomposition rule to synthesize arbitrary single-qubit operations using two $\rm \sqrt{X}$ gates interspersed with VZ gates \cite{McKay2017Efficient}. Consequently, it suffices to calibrate only the $\rm \sqrt{X}$ gate. It can be shown that any rotation $\rm R$, which maps the qubit from its ground state to the equator (XY plane) of the Bloch sphere, plus a virtual Z or phase rotation, can be equivalently used as a $\rm \sqrt{X}$ gate (see Supplemental Material \cite{SM} and Ref.~\cite{McKay2017Efficient}). Thus, calibrating such an $\rm R$ operation suffices to synthesize any single-qubit gate with the aid of VZ gates.

To calibrate the $\rm R$ gate for a given pulse shape, duration, carrier frequency, and DRAG parameters, we scan the drive amplitude until the excited-state population of the qubit (initialized in its ground state) reaches 0.5 (Fig.~\ref{fig: vz}(a)). We then concatenate two $\rm R$ gates with VZ gates to determine the necessary VZ phases for maximal excitation (Fig.\ref{fig: vz}(b)).
After fine-tuning these parameters using standard pulse techniques \cite{SM}, we validate the calibrated parameters $\Omega$ (drive amplitude) and $\phi$ (VZ phase) under intentional detunings, with results closely matching numerical simulations (Fig.~\ref{fig: vz}(c)). Using these VZ-compensated $\rm \sqrt{X}$ gates, we perform randomized benchmarking (RB) \cite{magesan_scalable_2011}. The errors per Clifford (EPC), shown in Fig.~\ref{fig: vz}(d), demonstrate consistent performance over a detuning range of $\pm 20$ MHz, showing that the VZ compensation method is suited for more general scenarios.

\begin{figure}[t!]
\centering
\includegraphics[scale=1]{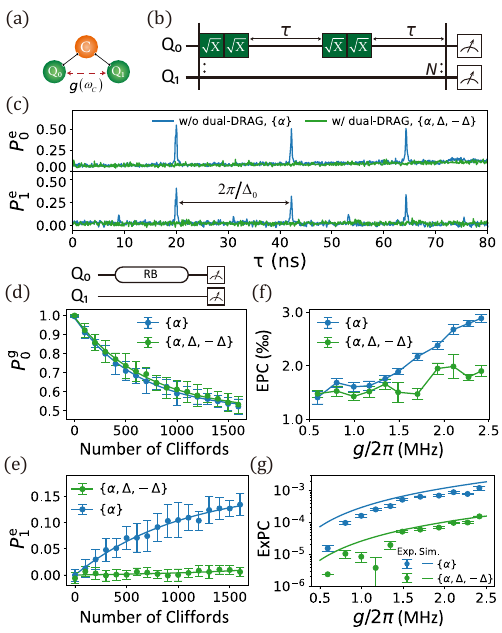}
\caption{\label{fig: suppression_qubit} 
(a) Experimental setup consisting of two transmon qubits and a tunable coupler, which is also a transmon qubit. The coupler is used to modulate the effective coupling strength between the qubits by adjusting its frequency.
(b) Pulse sequence designed for detecting spurious transitions in a nearby qubit during single-qubit gate operations. Repeated $\pi$ gates, each consisting of two $\pi/2$ pulses and spaced by a waiting time $\tau$, are applied to $\rm Q_0$.
(c) Measured excited-state populations of $\rm Q_0$ (top panel) and $\rm Q_1$ (bottom panel) using the detection sequence, as a function of the waiting time $\tau$. Results are shown with and without dual-DRAG. Both cases include DRAG correction at the leakage transition to the second excited state $\ket{2}$. The number of $\pi$ gate pairs is $N = 50$. Here, $g/2\pi = 1\,\rm{MHz}$, $\Delta_0/2\pi = 45\,\rm{MHz}$, and $t_g = 25\,\rm{ns}$.
(d) Randomized benchmarking results for $\rm Q_0$, comparing cases with and without dual-DRAG.
(e) Simultaneously monitored excited-state populations of $\rm Q_1$. The excitation rates per Clifford gate (ExPC) are $(0.8\pm0.3) \times 10^{-5}$ with dual-DRAG and $(1.6\pm0.2) \times 10^{-4}$ without dual-DRAG.
(f) Error per Clifford for different coupling strengths, with a fixed detuning of $\Delta_0/2\pi = 45~\rm{MHz}$.
(g) Excitation per Clifford for varying coupling strengths. The solid lines are the simulated ExPC averaged over 24 single-qubit Cliffords implemented with the U3 decomposition. The excitation rates are proportional to $g^2$ (see Supplemental Material \cite{SM} for details).}
\end{figure}

\emph{Suppressing spurious transitions between qubits} --- Next, we validate the dual-DRAG method's effectiveness in suppressing unwanted transitions due to quantum crosstalk. We use two frequency-tunable transmon qubits, with a tunable coupler (another transmon) adjusting the effective qubit-qubit coupling strength $g$ via the coupler frequency $\omega_{\rm c}$ (Fig.~\ref{fig: suppression_qubit}(a)). At first, we set the qubit-qubit detuning $\Delta_0 / 2\pi = 45\, \text{MHz}$ and the coupling $g / 2\pi = 1\, {\rm MHz}$.

For efficient detection of weak transitions ($g\Omega / \Delta_0 < \Delta_0$), we employ a protocol similar to the interference error filter \cite{lucero2008high,google2023suppressing} (Fig.~\ref{fig: suppression_qubit}(b)). An even number ($2N$) of $\rm X$ gates are applied to the control qubit $\rm Q_0$, keeping it in the ground state. 
Due to non-zero coupling, the spectator qubit $\rm Q_1$ is coherently excited by these pulses. A uniform waiting time $\tau$ is added after each $\pi$ gate, during which the wavefunction amplitude of $\rm Q_1$'s excited state (e.g., $\ket{\rm ge}$) accumulates a relative phase $e^{-\mathrm{i}\Delta_0\tau}$. At specific $\tau$, the transition amplitudes interfere constructively, amplifying $\rm Q_1$'s excitation.

Figure~\ref{fig: suppression_qubit}(c) shows the measured total populations of all excited states ($P^{\rm e}_{i}$, $i=0,1$), including $\ket{\rm e}$ and $\ket{\rm f}$, for both qubits as a function of $\tau$. Without dual-DRAG correction, strong periodic peaks appear on both qubits at intervals of $\sim22.2~\text{ns}$, matching $2\pi/\Delta_0$ and confirming our expectation. Since $\rm X$ gates are applied in pairs, transitions to the excited state of $\rm Q_1$ --- whether through $\ket{\rm gg}\to\ket{\rm ge}$ or $\ket{\rm eg}\to\ket{\rm ee}$ --- result in the $\ket{\rm ee}$ state in the end. Additionally, smaller secondary peaks on $\rm Q_1$, halfway between primary peaks and sharing the same periodicity, are attributed to residual $\rm IX$ interaction arising from both quantum and classical crosstalk \cite{SM}.

We repeat measurements with dual-DRAG correction at $\pm\Delta$. The parameter $\Delta$ is initially set to $\Delta_0$ and fine-tuned by minimizing excitation peaks. Using three DRAG parameters $\{\alpha, \Delta, -\Delta\}$, all prominent excitation peaks vanish (Fig.~\ref{fig: suppression_qubit}(c)), demonstrating effective suppression. Randomized benchmarking on $\rm Q_0$ (Fig.~\ref{fig: suppression_qubit}(d,e)) shows a slight error reduction (error per Clifford: $1.60 \times 10^{-3} \to 1.42 \times 10^{-3}$). Simultaneously, $\rm Q_1$'s excitation rate per Clifford (ExPC) drops by an order of magnitude from $(1.6 \pm 0.2) \times 10^{-4}$ to $(0.8 \pm 0.3) \times 10^{-5}$. Although the effect in EPC does not seem to be significant, the induced correlated excitations, in particular between distant qubits, complicate the error model and may severely affect error correction codes.

We further test the method across coupling strengths using the tunable coupler. The improvement in gate error rates using dual-DRAG becomes more pronounced at larger $g$ (Fig.~\ref{fig: suppression_qubit}(f)). This trend aligns with the measured spurious excitation rates, which scale quadratically with $g$ as expected from the cross-driving amplitude that scales with $g/\Delta_0$. At large $g$, these spurious excitations dominate the gate errors, yet dual-DRAG consistently suppresses them by an order of magnitude across the whole range (Fig.~\ref{fig: suppression_qubit}(g)). These results demonstrate the effectiveness of spectrally balanced DRAG in suppressing weakly detuned transitions and enhancing gate fidelities.

\begin{figure}[t!]
\includegraphics[scale=1]{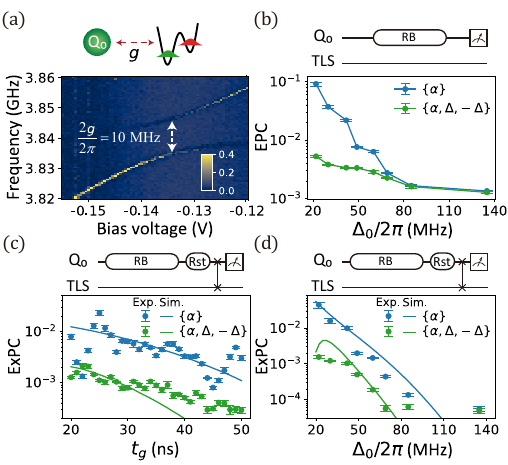}
\caption{\label{fig: tls}
(a) Spectroscopy of a tunable transmon qubit coupled to a spurious TLS, showing an avoided crossing of approximately 10\,MHz. The horizontal axis represents the voltage of the flux bias pulse applied to the coupler qubit.
(b) Error per Clifford of the qubit as a function of detuning between the qubit and the TLS. Here $t_g=25$\,ns.
(c) Excitation per Clifford of the TLS for different gate times at $\Delta_0/2\pi = 42\,\rm{MHz}$. The solid lines are twice the simulated $\rm \sqrt{X}$ gate errors.
(d) Excitation per Clifford of the TLS for varying detunings. The solid lines are twice the simulated $\rm \sqrt{X}$ gate errors.}
\end{figure}

\emph{Applications to the qubit-TLS system} --- We extend our method to suppress spurious transitions in systems where $\rm Q_0$ is coupled to a parasitic TLS. Such TLSs with long relaxation times are particularly problematic, as their unintended excitations can accumulate over time and persistently degrade subsequent gate operations. The specific TLS studied here exhibits a strong coupling ($g/2\pi = 5\,\text{MHz}$) to a transmon qubit, as evidenced by the avoided crossing in the measured qubit spectrum shown in Fig.~\ref{fig: tls}(a).

To characterize the dependence on detuning of our method, we calibrate the $\rm \sqrt{X}$ gate at varying qubit-TLS detunings $\Delta_0$ using the same protocol developed for coupled qubits. Figure~\ref{fig: tls}(b) shows RB results comparing dual-DRAG corrections (applied at $\pm\Delta$) to uncorrected pulses across $20\,\text{MHz} \leq \Delta_0/2\pi \leq 85\,\text{MHz}$. For $\Delta_0/2\pi < 20\,\text{MHz}$, spectral overlap between the drive frequency and the TLS transition hinders the convergence of parameters during calibration, and the dual-DRAG correction overcuts the spectral components around the target frequency, making the drive ineffective. At $\Delta_0/2\pi > 85\,\text{MHz}$, TLS-induced gate errors diminish to a negligible level.

We further quantify TLS excitation rates across gate time $t_g$ and detuning $\Delta_0$ (Fig.~\ref{fig: tls}(c,d)). To measure TLS excitations, we: (1) reset the qubit to its ground state post-RB, (2) apply an iSWAP gate to transfer TLS excitations to the qubit, and (3) perform qubit readout. In $20\,\text{ns} \leq t_g \leq 50\,\text{ns}$, dual-DRAG reduces TLS excitation rates by an order of magnitude compared to uncorrected pulses. Numerical simulations of single-pulse dynamics reproduce the observed decreasing trend, though deviations may arise from interference between pulses and spectral smoothing due to the presence of dephasing noise.

The measured excitation rates for $20\,\text{MHz} \leq \Delta_0/2\pi \leq 85\,\text{MHz}$ show a comparable improvement from the use of dual-DRAG. The rapidly enhanced suppression with increasing $\Delta_0$ results from the diminishing spectral weight at larger $\Delta_0$, and the reduced cross-drive amplitude that scales with $g/\Delta_0$.

\emph{Discussion} --- We have demonstrated a spectrally balanced pulse-shaping technique that suppresses spurious transitions in superconducting qubits. Our method reduces crosstalk errors by an order of magnitude and mitigates single-qubit gate errors from strongly coupled TLS. By engineering symmetric spectral holes, it effectively suppresses weakly detuned transitions by mitigating the off-resonance effect during pulses. The technique integrates seamlessly with virtual-Z gates, further improving gate fidelity.

This approach directly addresses key challenges in a frequency-crowded quantum processor: parasitic qubit couplings and TLS-induced errors, both of which hinder frequency allocation. By mitigating these effects, our work can enhance frequency planning flexibility and provide a framework for modeling microwave gate errors in complex environments. 

We note that while dual-DRAG pulses demonstrate strong performance for small-scale systems (as experimentally validated), their large-scale efficacy requires careful frequency arrangement to avoid extreme conditions with multiple near-resonant TLS (see Supplemental Material \cite{SM}). Future work should systematically investigate these scaling properties.
However, this scheme can be effectively combined with frequency-tuning approaches that isolate TLS from qubits \cite{2025-dane-ibm-performance,chen-2025-scalable,wang-escaping-2023} to further enhance gate performance, providing a comprehensive solution to the TLS problem.

The hardware-agnostic nature of the dual-DRAG protocol makes it broadly applicable to  other quantum architectures, including superconducting fluxonium qubits, semiconductor quantum dots, NV centers, and trapped ions, positioning it as a versatile tool for high-fidelity quantum control\cite{heinz2022,fang_crosstalk_2022,theis2016,Vezvaee2023Avoiding,takeda2018}.

\emph{Note added.} --- During the preparation of this manuscript, we became aware of a concurrent work \cite{matsuda2025selective} that independently develops a related pulse-shaping technique using symmetrically filtered spectra to address classical crosstalk.

\begin{acknowledgments}
The authors would like to thank Yu He, Yasunobu Nakamura and Peng Zhao for fruitful discussions. This work was supported by Beijing Natural Science Foundation (Grants No. JQ25014), National Natural Science Foundation of China (Grants No. 12404558, No. 12322413, No. 92365206, and No. 92476206), Innovation Program for Quantum Science and Technology (Grants No.2021ZD0301802).
\end{acknowledgments}

\emph{Data availability} --- The data that support the findings of this Letter are available from the authors upon reasonable request.

\bibliography{reference}

@article{kim2023evidence,
  title={Evidence for the utility of quantum computing before fault tolerance},
  url = {https://www.nature.com/articles/s41586-023-06096-3},
  author={Kim, Youngseok and Eddins, Andrew and Anand, Sajant and Wei, Ken Xuan and Van Den Berg, Ewout and Rosenblatt, Sami and Nayfeh, Hasan and Wu, Yantao and Zaletel, Michael and Temme, Kristan and Kandala, Abhinav},
  journal={Nature},
  volume={618},
  number={7965},
  pages={500--505},
  year={2023},
  publisher={Nature Publishing Group UK London}
}

@article{google2023suppressing,
  title={Suppressing quantum errors by scaling a surface code logical qubit},
  url = {https://www.nature.com/articles/s41586-022-05434-1},
  author = {Acharya, Rajeev and Aleiner, Igor and Allen, Richard and Andersen, Trond I. and Ansmann, Markus and Arute, Frank and Arya, Kunal and Asfaw, Abraham and Atalaya, Juan and Babbush, Ryan and Bacon, Dave and Bardin, Joseph C. and Basso, Joao and Bengtsson, Andreas and Boixo, Sergio and Bortoli, Gina and Bourassa, Alexandre and Bovaird, Jenna and Brill, Leon and Broughton, Michael and Buckley, Bob B. and Buell, David A. and Burger, Tim and Burkett, Brian and Bushnell, Nicholas and Chen, Yu and Chen, Zijun and Chiaro, Ben and Cogan, Josh and Collins, Roberto and Conner, Paul and Courtney, William and Crook, Alexander L. and Curtin, Ben and Debroy, Dripto M. and Del Toro Barba, Alexander and Demura, Sean and Dunsworth, Andrew and Eppens, Daniel and Erickson, Catherine and Faoro, Lara and Farhi, Edward and Fatemi, Reza and Flores Burgos, Leslie and Forati, Ebrahim and Fowler, Austin G. and Foxen, Brooks and Giang, William and Gidney, Craig and Gilboa, Dar and Giustina, Marissa and Grajales Dau, Alejandro and Gross, Jonathan A. and Habegger, Steve and Hamilton, Michael C. and Harrigan, Matthew P. and Harrington, Sean D. and Higgott, Oscar and Hilton, Jeremy and Hoffmann, Markus and Hong, Sabrina and Huang, Trent and Huff, Ashley and Huggins, William J. and Ioffe, Lev B. and Isakov, Sergei V. and Iveland, Justin and Jeffrey, Evan and Jiang, Zhang and Jones, Cody and Juhas, Pavol and Kafri, Dvir and Kechedzhi, Kostyantyn and Kelly, Julian and Khattar, Tanuj and Khezri, Mostafa and Kieferová, Mária and Kim, Seon and Kitaev, Alexei and Klimov, Paul V. and Klots, Andrey R. and Korotkov, Alexander N. and Kostritsa, Fedor and Kreikebaum, John Mark and Landhuis, David and Laptev, Pavel and Lau, Kim-Ming and Laws, Lily and Lee, Joonho and Lee, Kenny and Lester, Brian J. and Lill, Alexander and Liu, Wayne and Locharla, Aditya and Lucero, Erik and Malone, Fionn D. and Marshall, Jeffrey and Martin, Orion and McClean, Jarrod R. and McCourt, Trevor and McEwen, Matt and Megrant, Anthony and Meurer Costa, Bernardo and Mi, Xiao and Miao, Kevin C. and Mohseni, Masoud and Montazeri, Shirin and Morvan, Alexis and Mount, Emily and Mruczkiewicz, Wojciech and Naaman, Ofer and Neeley, Matthew and Neill, Charles and Nersisyan, Ani and Neven, Hartmut and Newman, Michael and Ng, Jiun How and Nguyen, Anthony and Nguyen, Murray and Niu, Murphy Yuezhen and O’Brien, Thomas E. and Opremcak, Alex and Platt, John and Petukhov, Andre and Potter, Rebecca and Pryadko, Leonid P. and Quintana, Chris and Roushan, Pedram and Rubin, Nicholas C. and Saei, Negar and Sank, Daniel and Sankaragomathi, Kannan and Satzinger, Kevin J. and Schurkus, Henry F. and Schuster, Christopher and Shearn, Michael J. and Shorter, Aaron and Shvarts, Vladimir and Skruzny, Jindra and Smelyanskiy, Vadim and Smith, W. Clarke and Sterling, George and Strain, Doug and Szalay, Marco and Torres, Alfredo and Vidal, Guifre and Villalonga, Benjamin and Vollgraff Heidweiller, Catherine and White, Theodore and Xing, Cheng and Yao, Z. Jamie and Yeh, Ping and Yoo, Juhwan and Young, Grayson and Zalcman, Adam and Zhang, Yaxing and Zhu, Ningfeng},
  journal={Nature},
  volume={614},
  number={7949},
  pages={676--681},
  year={2023},
  publisher={Nature Publishing Group UK London}
}

@article{cao2023generation,
  title={Generation of genuine entanglement up to 51 superconducting qubits},
  url = {https://www.nature.com/articles/s41586-023-06195-1},
  author={Cao, Sirui and Wu, Bujiao and Chen, Fusheng and Gong, Ming and Wu, Yulin and Ye, Yangsen and Zha, Chen and Qian, Haoran and Ying, Chong and Guo, Shaojun and Zhu, Qingling and Huang, He-Liang and Zhao, Youwei and Li, Shaowei and Wang, Shiyu and Yu, Jiale and Fan, Daojin and Wu, Dachao and Su, Hong and Deng, Hui and Rong, Hao and Li, Yuan and Zhang, Kaili and Chung, Tung-Hsun and Liang, Futian and Lin, Jin and Xu, Yu and Sun, Lihua and Guo, Cheng and Li, Na and Huo, Yong-Heng and Peng, Cheng-Zhi and Lu, Chao-Yang and Yuan, Xiao and Zhu, Xiaobo and Pan, Jian-Wei},
  journal={Nature},
  volume={619},
  number={7971},
  pages={738--742},
  year={2023},
  publisher={Nature Publishing Group UK London}
}

@article{xu2023digital,
  title={Digital simulation of projective non-Abelian anyons with 68 superconducting qubits},
  url = {https://iopscience.iop.org/article/10.1088/0256-307X/40/6/060301},
  author={Xu, Shibo and Sun, Zheng-Zhi and Wang, Ke and Xiang, Liang and Bao, Zehang and Zhu, Zitian and Shen, Fanhao and Song, Zixuan and Zhang, Pengfei and Ren, Wenhui and Zhang, Xu and Dong, Hang and Deng, Jinfeng and Chen, Jiachen and Wu, Yaozu and Tan, Ziqi and Gao, Yu and Jin, Feitong and Zhu, Xuhao and Zhang, Chuanyu and Wang, Ning and Zou, Yiren and Zhong, Jiarun and Zhang, Aosai and Li, Weikang and Jiang, Wenjie and Yu, Li-Wei and Yao, Yunyan and Wang, Zhen and Li, Hekang and Guo, Qiujiang and Song, Chao and Wang, H. and Deng, Dong-Ling},
  journal={Chinese Physics Letters},
  volume={40},
  number={6},
  pages={060301},
  year={2023},
  publisher={IOP Publishing}
}

@article{bluvstein2024logical,
  title={Logical quantum processor based on reconfigurable atom arrays},
  url = {https://www.nature.com/articles/s41586-023-06927-3},
  author={Bluvstein, Dolev and Evered, Simon J. and Geim, Alexandra A. and Li, Sophie H. and Zhou, Hengyun and Manovitz, Tom and Ebadi, Sepehr and Cain, Madelyn and Kalinowski, Marcin and Hangleiter, Dominik and Bonilla Ataides, J. Pablo and Maskara, Nishad and Cong, Iris and Gao, Xun and Sales Rodriguez, Pedro and Karolyshyn, Thomas and Semeghini, Giulia and Gullans, Michael J. and Greiner, Markus and Vuletić, Vladan and Lukin, Mikhail D.},
  journal={Nature},
  volume={626},
  number={7997},
  pages={58--65},
  year={2024},
  publisher={Nature Publishing Group UK London}
}

@article{Sarovar2020Detecting,
  title={Detecting crosstalk errors in quantum information processors},
  url = {http://arxiv.org/abs/1908.09855},
  author={Sarovar, Mohan and Proctor, Timothy and Rudinger, Kenneth and Young, Kevin and Nielsen, Erik and Blume-Kohout, Robin},
  journal={Quantum},
  volume={4},
  pages={321},
  year={2020},
  publisher={Verein zur F{\"o}rderung des Open Access Publizierens in den Quantenwissenschaften}
}

@article{Motzoi2009Simple,
  title={Simple pulses for elimination of leakage in weakly nonlinear qubits},
  url = {https://link.aps.org/doi/10.1103/PhysRevLett.103.110501},
  author={Motzoi, Felix and Gambetta, Jay M and Rebentrost, Patrick and Wilhelm, Frank K},
  journal={Physical Review Letters},
  volume={103},
  number={11},
  pages={110501},
  year={2009},
  publisher={APS}
}

@article{motzoi2013improving,
  title={Improving frequency selection of driven pulses using derivative-based transition suppression},
  url = {https://link.aps.org/doi/10.1103/PhysRevA.88.062318},
  author={Motzoi, Felix and Wilhelm, Frank K},
  journal={Physical Review A—Atomic, Molecular, and Optical Physics},
  volume={88},
  number={6},
  pages={062318},
  year={2013},
  publisher={APS}
}

@article{Chen2016Measuring,
  title={Measuring and suppressing quantum state leakage in a superconducting qubit},
  url = {https://link.aps.org/doi/10.1103/PhysRevLett.116.020501},
  author={Chen, Zijun and Kelly, Julian and Quintana, Chris and Barends, R. and Campbell, B. and Chen, Yu and Chiaro, B. and Dunsworth, A. and Fowler, A. G. and Lucero, E. and Jeffrey, E. and Megrant, A. and Mutus, J. and Neeley, M. and Neill, C. and O’Malley, P. J. J. and Roushan, P. and Sank, D. and Vainsencher, A. and Wenner, J. and White, T. C. and Korotkov, A. N. and Martinis, John M.},
  journal={Physical Review Letters},
  volume={116},
  number={2},
  pages={020501},
  year={2016},
  publisher={APS}
}

@article{Gambetta2011Analytic,
  title={Analytic control methods for high-fidelity unitary operations in a weakly nonlinear oscillator},
  url = {https://link.aps.org/doi/10.1103/PhysRevA.83.012308},
  author={Gambetta, Jay M and Motzoi, F and Merkel, ST and Wilhelm, Frank K},
  journal={Physical Review A—Atomic, Molecular, and Optical Physics},
  volume={83},
  number={1},
  pages={012308},
  year={2011},
  publisher={APS}
}

@article{Vezvaee2023Avoiding,
  title={Avoiding leakage and errors caused by unwanted transitions in lambda systems},
  url = {https://link.aps.org/doi/10.1103/PRXQuantum.4.030312},
  author={Vezvaee, Arian and Takou, Evangelia and Hilaire, Paul and Doty, Matthew F and Economou, Sophia E},
  journal={PRX Quantum},
  volume={4},
  number={3},
  pages={030312},
  year={2023},
  publisher={APS}
}

@article{Malekakhlagh2022Mitigating,
  title={Mitigating off-resonant error in the cross-resonance gate},
  url = {https://link.aps.org/doi/10.1103/PhysRevA.105.012602},
  author={Malekakhlagh, Moein and Magesan, Easwar},
  journal={Physical Review A},
  volume={105},
  number={1},
  pages={012602},
  year={2022},
  publisher={APS}
}

@article{Wei2024Characterizing,
  title={Characterizing non-Markovian off-resonant errors in quantum gates},
  url = {https://doi.org/10.1103/PhysRevApplied.21.024018},
  author={Wei, Ken Xuan and Pritchett, Emily and Zajac, David M and McKay, David C and Merkel, Seth},
  journal={Physical Review Applied},
  volume={21},
  number={2},
  pages={024018},
  year={2024},
  publisher={APS}
}

@article{Li2023Suppression,
  title={Experimental error suppression in Cross-Resonance gates via multi-derivative pulse shaping},
  url = {https://www.nature.com/articles/s41534-024-00863-4},
  author={Li, Boxi and Calarco, Tommaso and Motzoi, Felix},
  journal={npj Quantum Information},
  volume={10},
  number={1},
  pages={66},
  year={2024},
  publisher={Nature Publishing Group UK London}
}

@article{Theis2016Simultaneous,
  title={Simultaneous gates in frequency-crowded multilevel systems using fast, robust, analytic control shapes},
  url = {https://link.aps.org/doi/10.1103/PhysRevA.93.012324},
  author={Theis, LS and Motzoi, F and Wilhelm, FK},
  journal={Physical Review A},
  volume={93},
  number={1},
  pages={012324},
  year={2016},
  publisher={APS}
}

@article{Nuerbolati2022Canceling,
  title={Canceling microwave crosstalk with fixed-frequency qubits},
  url = {https://aip.scitation.org/doi/10.1063/5.0088094},
  author={Nuerbolati, Wuerkaixi and Han, Zhikun and Chu, Ji and Zhou, Yuxuan and Tan, Xinsheng and Yu, Yang and Liu, Song and Yan, Fei},
  journal={Applied Physics Letters},
  volume={120},
  number={17},
  year={2022},
  publisher={AIP Publishing}
}

@article{rigetti2010fully,
  title={Fully microwave-tunable universal gates in superconducting qubits with linear couplings and fixed transition frequencies},
  url = {https://link.aps.org/doi/10.1103/PhysRevB.81.134507},
  author={Rigetti, Chad and Devoret, Michel},
  journal={Physical Review B—Condensed Matter and Materials Physics},
  volume={81},
  number={13},
  pages={134507},
  year={2010},
  publisher={APS}
}

@article{de2010selective,
  title={Selective darkening of degenerate transitions demonstrated with two superconducting quantum bits},
  url = {https://www.nature.com/articles/nphys1733},
  author={De Groot, PC and Lisenfeld, J and Schouten, RN and Ashhab, S and Lupa{\c{s}}cu, A and Harmans, CJPM and Mooij, JE},
  journal={Nature Physics},
  volume={6},
  number={10},
  pages={763--766},
  year={2010},
  publisher={Nature Publishing Group UK London}
}

@article{Tripathi2019Operation,
  title={Operation and intrinsic error budget of a two-qubit cross-resonance gate},
  url = {https://link.aps.org/doi/10.1103/PhysRevA.100.012301},
  author={Tripathi, Vinay and Khezri, Mostafa and Korotkov, Alexander N},
  journal={Physical Review A},
  volume={100},
  number={1},
  pages={012301},
  year={2019},
  publisher={APS}
}

@unpublished{SM,
	title  = {},
	author = {},
	year   = {},
	note   = {See Supplemental Material at [URL] for the supporting information of the experiments and simulations.}
}

@article{Mckay2017Efficient,
  title={Efficient Z gates for quantum computing},
  url = {https://link.aps.org/doi/10.1103/PhysRevA.96.022330},
  author={McKay, David C and Wood, Christopher J and Sheldon, Sarah and Chow, Jerry M and Gambetta, Jay M},
  journal={Physical Review A},
  volume={96},
  number={2},
  pages={022330},
  year={2017},
  publisher={APS}
}

@article{matsuda2025selective,
  title={Selective Excitation of Superconducting Qubits with a Shared Control Line through Pulse Shaping},
  url = {http://arxiv.org/abs/2501.10710},
  author={Matsuda, R. and Ohira, R. and Sumida, T. and Shiomi, H. and Machino, A. and Morisaka, S. and Koike, K. and Miyoshi, T. and Kurimoto, Y. and Sugita, Y. and Ito, Y. and Suzuki, Y. and Spring, P. A. and Wang, S. and Tamate, S. and Tabuchi, Y. and Nakamura, Y. and Ogawa, K. and Negoro, M.},
  journal={arXiv preprint arXiv:2501.10710},
  year={2025}
}

@article{muller_towards_2019,
	title = {Towards understanding two-level-systems in amorphous solids: insights from quantum circuits},
	volume = {82},
	issn = {0034-4885, 1361-6633},
	shorttitle = {Towards understanding two-level-systems in amorphous solids},
	url = {https://iopscience.iop.org/article/10.1088/1361-6633/ab3a7e},
	doi = {10.1088/1361-6633/ab3a7e},
	number = {12},
	urldate = {2022-09-09},
	journal = {Reports on Progress in Physics},
	author = {Müller, Clemens and Cole, Jared H and Lisenfeld, Jürgen},
	month = dec,
	year = {2019},
	pages = {124501},
}

@article{klimov_optimizing_2024,
	title = {Optimizing quantum gates towards the scale of logical qubits},
	volume = {15},
	copyright = {2024 The Author(s)},
	issn = {2041-1723},
	url = {https://www.nature.com/articles/s41467-024-46623-y},
	doi = {10.1038/s41467-024-46623-y},
	number = {1},
	urldate = {2025-01-02},
	journal = {Nature Communications},
	author = {Klimov, Paul V. and Bengtsson, Andreas and Quintana, Chris and Bourassa, Alexandre and Hong, Sabrina and Dunsworth, Andrew and Satzinger, Kevin J. and Livingston, William P. and Sivak, Volodymyr and Niu, Murphy Yuezhen and Andersen, Trond I. and Zhang, Yaxing and Chik, Desmond and Chen, Zijun and Neill, Charles and Erickson, Catherine and Grajales Dau, Alejandro and Megrant, Anthony and Roushan, Pedram and Korotkov, Alexander N. and Kelly, Julian and Smelyanskiy, Vadim and Chen, Yu and Neven, Hartmut},
	month = mar,
	year = {2024},
	note = {Publisher: Nature Publishing Group},
	pages = {2442},
}

@article{bylander_noise_2011,
	title = {Noise spectroscopy through dynamical decoupling with a superconducting flux qubit},
	volume = {7},
	copyright = {2011 Springer Nature Limited},
	issn = {1745-2481},
	url = {https://www.nature.com/articles/nphys1994},
	doi = {10.1038/nphys1994},
	number = {7},
	urldate = {2024-01-19},
	journal = {Nature Physics},
	author = {Bylander, Jonas and Gustavsson, Simon and Yan, Fei and Yoshihara, Fumiki and Harrabi, Khalil and Fitch, George and Cory, David G. and Nakamura, Yasunobu and Tsai, Jaw-Shen and Oliver, William D.},
	month = jul,
	year = {2011},
	note = {Number: 7
Publisher: Nature Publishing Group},
	pages = {565--570},
}

@article{chen2025efficient,
  title={Efficient Implementation of Arbitrary Two-Qubit Gates via Unified Control},
  url = {https://doi.org/10.1038/s41567-025-02990-x},
  author={Chen, Zhen and Liu, Weiyang and Ma, Yanjun and Sun, Weijie and Wang, Ruixia and Wang, He and Xu, Huikai and Xue, Guangming and Yan, Haisheng and Yang, Zhen and Ding, Jiayu and Gao, Yang and Li, Feiyu and Zhang, Yujia and Zhang, Zikang and Jin, Yirong and Yu, Haifeng and Chen, Jianxin and Yan, Fei},
  journal={Natrue Physics },
  year={2025},
  volume={21},
  pages={1489-–1496}
}

@article{niu2019learning,
  title={Learning Non-Markovian Quantum Noise from Moir\'{e}-Enhanced Swap Spectroscopy with Deep Evolutionary Algorithm},
  author={Niu, Murphy Yuezhen and Smelyanskyi, Vadim and Klimov, Paul and Boixo, Sergio and Barends, Rami and Kelly, Julian and Chen, Yu and Arya, Kunal and Burkett, Brian and Bacon, Dave and others},
  journal={arXiv preprint arXiv:1912.04368},
  year={2019}
}

@article{magesan_scalable_2011,
	title = {Scalable and {Robust} {Randomized} {Benchmarking} of {Quantum} {Processes}},
	volume = {106},
	url = {https://link.aps.org/doi/10.1103/PhysRevLett.106.180504},
	doi = {10.1103/PhysRevLett.106.180504},
	number = {18},
	urldate = {2025-02-10},
	journal = {Physical Review Letters},
	author = {Magesan, Easwar and Gambetta, J. M. and Emerson, Joseph},
	month = may,
	year = {2011},
	note = {Publisher: American Physical Society},
	pages = {180504},
}

@article{acharya_quantum_2024,
title = {Quantum error correction below the surface code threshold},
copyright = {2024 The Author(s)},
issn = {1476-4687},
url = {https://www.nature.com/articles/s41586-024-08449-y},
doi = {10.1038/s41586-024-08449-y},
urldate = {2025-02-10},
journal = {Nature},
author = {Acharya, Rajeev and Abanin, Dmitry A. and Aghababaie-Beni, Laleh and Aleiner, Igor and Andersen, Trond I. and Ansmann, Markus and Arute, Frank and Arya, Kunal and Asfaw, Abraham and Astrakhantsev, Nikita and Atalaya, Juan and Babbush, Ryan and Bacon, Dave and Ballard, Brian and Bardin, Joseph C. and Bausch, Johannes and Bengtsson, Andreas and Bilmes, Alexander and Blackwell, Sam and Boixo, Sergio and Bortoli, Gina and Bourassa, Alexandre and Bovaird, Jenna and Brill, Leon and Broughton, Michael and Browne, David A. and Buchea, Brett and Buckley, Bob B. and Buell, David A. and Burger, Tim and Burkett, Brian and Bushnell, Nicholas and Cabrera, Anthony and Campero, Juan and Chang, Hung-Shen and Chen, Yu and Chen, Zijun and Chiaro, Ben and Chik, Desmond and Chou, Charina and Claes, Jahan and Cleland, Agnetta Y. and Cogan, Josh and Collins, Roberto and Conner, Paul and Courtney, William and Crook, Alexander L. and Curtin, Ben and Das, Sayan and Davies, Alex and De Lorenzo, Laura and Debroy, Dripto M. and Demura, Sean and Devoret, Michel and Di Paolo, Agustin and Donohoe, Paul and Drozdov, Ilya and Dunsworth, Andrew and Earle, Clint and Edlich, Thomas and Eickbusch, Alec and Elbag, Aviv Moshe and Elzouka, Mahmoud and Erickson, Catherine and Faoro, Lara and Farhi, Edward and Ferreira, Vinicius S. and Burgos, Leslie Flores and Forati, Ebrahim and Fowler, Austin G. and Foxen, Brooks and Ganjam, Suhas and Garcia, Gonzalo and Gasca, Robert and Genois, Elie and Giang, William and Gidney, Craig and Gilboa, Dar and Gosula, Raja and Dau, Alejandro Grajales and Graumann, Dietrich and Greene, Alex and Gross, Jonathan A. and Habegger, Steve and Hall, John and Hamilton, Michael C. and Hansen, Monica and Harrigan, Matthew P. and Harrington, Sean D. and Heras, Francisco J. H. and Heslin, Stephen and Heu, Paula and Higgott, Oscar and Hill, Gordon and Hilton, Jeremy and Holland, George and Hong, Sabrina and Huang, Hsin-Yuan and Huff, Ashley and Huggins, William J. and Ioffe, Lev B. and Isakov, Sergei V. and Iveland, Justin and Jeffrey, Evan and Jiang, Zhang and Jones, Cody and Jordan, Stephen and Joshi, Chaitali and Juhas, Pavol and Kafri, Dvir and Kang, Hui and Karamlou, Amir H. and Kechedzhi, Kostyantyn and Kelly, Julian and Khaire, Trupti and Khattar, Tanuj and Khezri, Mostafa and Kim, Seon and Klimov, Paul V. and Klots, Andrey R. and Kobrin, Bryce and Kohli, Pushmeet and Korotkov, Alexander N. and Kostritsa, Fedor and Kothari, Robin and Kozlovskii, Borislav and Kreikebaum, John Mark and Kurilovich, Vladislav D. and Lacroix, Nathan and Landhuis, David and Lange-Dei, Tiano and Langley, Brandon W. and Laptev, Pavel and Lau, Kim-Ming and Le Guevel, Loïck and Ledford, Justin and Lee, Joonho and Lee, Kenny and Lensky, Yuri D. and Leon, Shannon and Lester, Brian J. and Li, Wing Yan and Li, Yin and Lill, Alexander T. and Liu, Wayne and Livingston, William P. and Locharla, Aditya and Lucero, Erik and Lundahl, Daniel and Lunt, Aaron and Madhuk, Sid and Malone, Fionn D. and Maloney, Ashley and Mandrà, Salvatore and Manyika, James and Martin, Leigh S. and Martin, Orion and Martin, Steven and Maxfield, Cameron and McClean, Jarrod R. and McEwen, Matt and Meeks, Seneca and Megrant, Anthony and Mi, Xiao and Miao, Kevin C. and Mieszala, Amanda and Molavi, Reza and Molina, Sebastian and Montazeri, Shirin and Morvan, Alexis and Movassagh, Ramis and Mruczkiewicz, Wojciech and Naaman, Ofer and Neeley, Matthew and Neill, Charles and Nersisyan, Ani and Neven, Hartmut and Newman, Michael and Ng, Jiun How and Nguyen, Anthony and Nguyen, Murray and Ni, Chia-Hung and Niu, Murphy Yuezhen and O’Brien, Thomas E. and Oliver, William D. and Opremcak, Alex and Ottosson, Kristoffer and Petukhov, Andre and Pizzuto, Alex and Platt, John and Potter, Rebecca and Pritchard, Orion and Pryadko, Leonid P. and Quintana, Chris and Ramachandran, Ganesh and Reagor, Matthew J. and Redding, John and Rhodes, David M. and Roberts, Gabrielle and Rosenberg, Eliott and Rosenfeld, Emma and Roushan, Pedram and Rubin, Nicholas C. and Saei, Negar and Sank, Daniel and Sankaragomathi, Kannan and Satzinger, Kevin J. and Schurkus, Henry F. and Schuster, Christopher and Senior, Andrew W. and Shearn, Michael J. and Shorter, Aaron and Shutty, Noah and Shvarts, Vladimir and Singh, Shraddha and Sivak, Volodymyr and Skruzny, Jindra and Small, Spencer and Smelyanskiy, Vadim and Smith, W. Clarke and Somma, Rolando D. and Springer, Sofia and Sterling, George and Strain, Doug and Suchard, Jordan and Szasz, Aaron and Sztein, Alex and Thor, Douglas and Torres, Alfredo and Torunbalci, M. Mert and Vaishnav, Abeer and Vargas, Justin and Vdovichev, Sergey and Vidal, Guifre and Villalonga, Benjamin and Heidweiller, Catherine Vollgraff and Waltman, Steven and Wang, Shannon X. and Ware, Brayden and Weber, Kate and Weidel, Travis and White, Theodore and Wong, Kristi and Woo, Bryan W. K. and Xing, Cheng and Yao, Z. Jamie and Yeh, Ping and Ying, Bicheng and Yoo, Juhwan and Yosri, Noureldin and Young, Grayson and Zalcman, Adam and Zhang, Yaxing and Zhu, Ningfeng and Zobrist, Nicholas},
month = dec,
year = {2024},
note = {Publisher: Nature Publishing Group},
pages = {1--7},
}

@article{ghosh_understanding_2013,
	title = {Understanding the effects of leakage in superconducting quantum-error-detection circuits},
	volume = {88},
	url = {https://link.aps.org/doi/10.1103/PhysRevA.88.062329},
	doi = {10.1103/PhysRevA.88.062329},
	number = {6},
	urldate = {2025-02-11},
	journal = {Physical Review A},
	author = {Ghosh, Joydip and Fowler, Austin G. and Martinis, John M. and Geller, Michael R.},
	month = dec,
	year = {2013},
	note = {Publisher: American Physical Society},
	pages = {062329},
}

@article{iqbal_non-abelian_2024,
title = {Non-Abelian topological order and anyons on a trapped-ion processor},
volume = {626},
copyright = {2024 The Author(s), under exclusive licence to Springer Nature Limited},
issn = {1476-4687},
url = {https://www.nature.com/articles/s41586-023-06934-4},
doi = {10.1038/s41586-023-06934-4},
number = {7999},
urldate = {2025-02-13},
journal = {Nature},
author = {Iqbal, Mohsin and Tantivasadakarn, Nathanan and Verresen, Ruben and Campbell, Sara L. and Dreiling, Joan M. and Figgatt, Caroline and Gaebler, John P. and Johansen, Jacob and Mills, Michael and Moses, Steven A. and Pino, Juan M. and Ransford, Anthony and Rowe, Mary and Siegfried, Peter and Stutz, Russell P. and Foss-Feig, Michael and Vishwanath, Ashvin and Dreyer, Henrik},
month = feb,
year = {2024},
note = {Publisher: Nature Publishing Group},
pages = {505--511},
}

@inproceedings{ding2020systematic,
  title={Systematic crosstalk mitigation for superconducting qubits via frequency-aware compilation},
  url = {
https://doi.org/10.1109/MICRO50266.2020.00028},
  author={Ding, Yongshan and Gokhale, Pranav and Lin, Sophia Fuhui and Rines, Richard and Propson, Thomas and Chong, Frederic T},
  booktitle={2020 53rd Annual IEEE/ACM International Symposium on Microarchitecture (MICRO)},
  pages={201--214},
  year={2020},
  organization={IEEE}
}

@article{zhao2023mitigation,
  title={Mitigation of quantum crosstalk in cross-resonance-based qubit architectures},
  url = {https://doi.org/10.1103/PhysRevApplied.20.054033},
  author={Zhao, Peng},
  journal={Physical Review Applied},
  volume={20},
  number={5},
  pages={054033},
  year={2023},
  publisher={APS}
}

@inproceedings{murali2020software,
  title={Software mitigation of crosstalk on noisy intermediate-scale quantum computers},
  url = {
https://doi.org/10.1145/3373376.3378477},
  author={Murali, Prakash and McKay, David C and Martonosi, Margaret and Javadi-Abhari, Ali},
  booktitle={Proceedings of the Twenty-Fifth International Conference on Architectural Support for Programming Languages and Operating Systems},
  pages={1001--1016},
  year={2020}
}

@article{fang_crosstalk_2022,
	title = {Crosstalk {Suppression} in {Individually} {Addressed} {Two}-{Qubit} {Gates} in a {Trapped}-{Ion} {Quantum} {Computer}},
	volume = {129},
	issn = {0031-9007, 1079-7114},
	url = {https://link.aps.org/doi/10.1103/PhysRevLett.129.240504},
	doi = {10.1103/PhysRevLett.129.240504},
	number = {24},
	urldate = {2025-04-15},
	journal = {Physical Review Letters},
	author = {Fang, Chao and Wang, Ye and Huang, Shilin and Brown, Kenneth R. and Kim, Jungsang},
	month = dec,
	year = {2022},
	pages = {240504}
}

@article{theis2016,
title = {High fidelity Rydberg-blockade entangling gate using shaped, analytic pulses},
volume = {94},
copyright = {http://link.aps.org/licenses/aps-default-license},
issn = {2469-9926, 2469-9934},
url = {https://link.aps.org/doi/10.1103/PhysRevA.94.032306},
doi = {10.1103/PhysRevA.94.032306},
number = {3},
urldate = {2025-04-15},
journal = {Physical Review A},
author = {Theis, L. S. and Motzoi, F. and Wilhelm, F. K. and Saffman, M.},
month = sep,
year = {2016},
pages = {032306}
}

@article{takeda2018,
title = {{Optimized} electrical control of a {Si}/{SiGe} spin qubit in the presence of an induced frequency shift},
volume = {4},
issn = {2056-6387},
url = {https://www.nature.com/articles/s41534-018-0105-z},
doi = {10.1038/s41534-018-0105-z},
number = {1},
urldate = {2025-04-15},
journal = {npj Quantum Information},
author = {Takeda, K. and Yoneda, J. and Otsuka, T. and Nakajima, T. and Delbecq, M. R. and Allison, G. and Hoshi, Y. and Usami, N. and Itoh, K. M. and Oda, S. and Kodera, T. and Tarucha, S.},
month = oct,
year = {2018},
pages = {54}
}

@article{heinz2022,
	title = {{Crosstalk} analysis for simultaneously driven two-qubit gates in spin qubit arrays},
	volume = {105},
	issn = {2469-9950, 2469-9969},
	url = {https://link.aps.org/doi/10.1103/PhysRevB.105.085414},
	doi = {10.1103/PhysRevB.105.085414},
	number = {8},
	urldate = {2025-04-16},
	journal = {Physical Review B},
	author = {Heinz, Irina and Burkard, Guido},
	month = feb,
	year = {2022},
	pages = {085414}
}

@article{hyyppa2024reducing,
title = {{Reducing} {Leakage} of {Single}-{Qubit} {Gates} for {Superconducting} {Quantum} {Processors} {Using} {Analytical} {Control} {Pulse} {Envelopes}},
volume = {5},
issn = {2691-3399},
url = {https://link.aps.org/doi/10.1103/PRXQuantum.5.030353},
doi = {10.1103/PRXQuantum.5.030353},
number = {3},
urldate = {2025-04-05},
journal = {PRX Quantum},
author = {Hyyppä, Eric and Vepsäläinen, Antti and Papič, Miha and Chan, Chun Fai and Inel, Sinan and Landra, Alessandro and Liu, Wei and Luus, Jürgen and Marxer, Fabian and Ockeloen-Korppi, Caspar and Orbell, Sebastian and Tarasinski, Brian and Heinsoo, Johannes},
month = sep,
year = {2024},
pages = {030353}
}

@article{lucero2008high,
	title = {{High}-{Fidelity} {Gates} in a {Single} {Josephson} {Qubit}},
	volume = {100},
	copyright = {http://link.aps.org/licenses/aps-default-license},
	issn = {0031-9007, 1079-7114},
	url = {https://link.aps.org/doi/10.1103/PhysRevLett.100.247001},
	doi = {10.1103/PhysRevLett.100.247001},
	number = {24},
	urldate = {2025-04-19},
	journal = {Physical Review Letters},
	author = {Lucero, Erik and Hofheinz, M. and Ansmann, M. and Bialczak, Radoslaw C. and Katz, N. and Neeley, Matthew and O’Connell, A. D. and Wang, H. and Cleland, A. N. and Martinis, John M.},
	month = jun,
	year = {2008},
	pages = {247001}
}

@article{2025-dane-ibm-performance,
title = {{Performance} {Stabilization} of {High}-{Coherence} {Superconducting} {Qubits}},
url = {http://arxiv.org/abs/2503.12514},
urldate = {2025-07-16},
publisher = {arXiv},
author = {Dane, Andrew and Balakrishnan, Karthik and Wacaser, Brent and Hung, Li-Wen and Mamin, H. J. and Rugar, Daniel and Shelby, Robert M. and Murray, Conal and Rodbell, Kenneth and Sleight, Jeffrey},
month = mar,
year = {2025},
journal = {arXiv preprint arXiv:2503.12514}
}

@article{chen-2025-scalable,
title = {{Scalable} and {Site}-{Specific} {Frequency} {Tuning} of {Two}-{Level} {System} {Defects} in {Superconducting} {Qubit} {Arrays}},
url = {http://arxiv.org/abs/2503.04702},
urldate = {2025-03-07},
journal = {arXiv preprint arXiv:2503.04702},
author = {Chen, Larry and Lee, Kan-Heng and Liu, Chuan-Hong and Marinelli, Brian and Naik, Ravi K. and Kang, Ziqi and Goss, Noah and Kim, Hyunseong and Santiago, David I. and Siddiqi, Irfan},
year = {2025}
}

@article{wang-escaping-2023,
title = {Escaping {Detrimental} {Interactions} with {Microwave}-{Dressed} {Transmon} {Qubits}},
volume = {40},
issn = {0256-307X, 1741-3540},
url = {https://iopscience.iop.org/article/10.1088/0256-307X/40/7/070304},
doi = {10.1088/0256-307X/40/7/070304},
number = {7},
urldate = {2023-11-09},
journal = {Chinese Physics Letters},
author = {Wang, Z. T. and Zhao, Peng and Yang, Z. H. and Tian, Ye and Yu, H. F. and Zhao, S. P.},
month = jul,
year = {2023},
pages = {070304}
}

\newpage   
\clearpage 
\onecolumngrid

\begin{center}
\textbf{\large Supplemental Materials for ``Suppressing spurious transitions using spectrally balanced pulse"}
\end{center}

\section{Experimental setup}\label{append:exp_setup}

The superconducting quantum processor used in this study and the experimental setup closely mirror those described in Ref.~\cite{chen2025efficient}. Qubits $\rm Q_0$ and $\rm Q_1$ are frequency-tunable transmon qubits, coupled through a tunable coupler. The coupling strength between the qubits can be adjusted by tuning the coupler’s frequency. Qubit $\rm Q_0$ is also employed in the TLS experiment and is coupled to the relevant TLS. Table \ref{tab:table1} lists the maximum ($\omega^{\rm max}$) and typical operating ($\omega^{\rm idle}$) frequencies of the qubits and the TLS $\ket{0} \rightarrow \ket{1}$ transition, as well as the anharmonicities of the qubits. Additionally, it provides the $T_1$ relaxation times measured at both maximum and operating frequencies, and the $T_2$ times obtained via Ramsey interferometry at those frequencies.

\begin{table}[h!]
\begin{threeparttable}
    \centering
    \label{tab:table1}
    \begin{tabular}{|c|c|c|c|c|}
    \hline
        & $\rm Q_0$ & $\rm Q_1$ & TLS \\
        \colrule
        ~~$\omega^{\rm max}/2\pi$ (GHz)~~ & ~~3.91~~ & ~~3.76~~ & ~3.84~~\\

        ~~$\omega^{\rm idle}/2\pi$ (GHz)~~ & ~~3.76~~ & ~~3.72~~ & ~3.84~~\\
        
        ~~$\alpha/2\pi$ (MHz) ~~& ~~-194.6~~ & ~~-193.2~~ & $-$\\
       
        ~~$T_1^{\rm max}$ ($\mu$s) ~~ & ~~78.6~~ & ~~72.6~~ & $-$\\

        ~~$T_1^{\rm idle}$ ($\mu$s) ~~ & ~~78.1~~ & ~~76.8~~ & ~14.32~~\\
        
        ~~$T_{2R}^{\rm max}$ ($\mu$s) ~~ & ~~37.3~~ & ~~22.2~~ & $-$ \\
        
        ~~$T_{2R}^{\rm idle}$ ($\mu$s) ~~ & ~~1.1~~ & ~~5.5~~ & $-$ \\
    \hline
    \end{tabular}
\end{threeparttable}
\caption{\label{app: table_2q} Parameters describing the two qubits and qubit-TLS system studied in our experiment.}
\end{table}

\section{Theoretical model}\label{append:model}
The Hamiltonian for the two-qubit system is ($\hbar = 1$)

\begin{equation}
\label{Equ: Hamiltonian}
\begin{split}
H = &\ H_0 + H_D,\\
H_0 = &\ \sum_{i=0,1}(\omega_i a_i^{\dagger}a_i + \frac{\alpha_i}{2}a_i^{\dagger} a_i^{\dagger}a_i a_i) +g(a_0^{\dagger}a_1+H.c.),\\
H_D = &\ \Omega(t)(a_0^{\dagger}+a_0).
\end{split}
\end{equation}

$H_0$ is the drift Hamiltonian, with a constant coupling strength $g$ between the two qubits, while $H_D$ is the driving Hamiltonian, representing the single-qubit operation on $\rm Q_0$. In the absence of a driving pulse, the dressed states of the Hamiltonian $H_0$ are denoted as $|{\rm ij}\rangle$, where the eigenfrequencies are given by $\Lambda_{{\rm ij}}$ ($\rm i,j = g,e$). The driving frequency applied to $\rm Q_0$ is set to $\Lambda_{\rm eg}$. In the dressed state basis ${|{\rm gg}\rangle, |{\rm ge}\rangle, |{\rm eg}\rangle, |{\rm ee}\rangle}$, and using the rotating wave approximation along with the first-order approximation, we have

\begin{eqnarray}
H_r = \left[\begin{array}{cccc}
 0 & \frac{\Omega g}{2\Delta_0} & \frac{\Omega}{2} & 0\\
 \frac{\Omega g}{2\Delta_0} & \delta+\Delta_0 & 0 & \frac{\Omega}{2}\\
 \frac{\Omega}{2} & 0 & \delta & -\frac{\Omega g}{2\Delta_0} \\
 0 & \frac{\Omega}{2} & -\frac{\Omega g}{2\Delta_0} & 2\delta+\Delta_0
\end{array}\right],
\label{eqM}
\end{eqnarray}
where $\Delta_0 = \Lambda_{\rm ge} - \Lambda_{\rm eg}$, $\delta = \Lambda_{\rm eg}-\omega_{{\rm d}}$ and $\omega_{{\rm d}}$ is the driving frequency.

For simplicity, in the Hamiltonian model and DRAG correction analysis (Section \ref{sec:app-DRAG}), we only consider the ideal two-level system. 
But for the transmon qubits, impacted by the higher energy levels, the Hamiltonian of the quantum crosstalk has $\rm ZX$ and $\rm IX$ components \cite{Tripathi2019Operation}. So that, in Section \ref{sec: app-Detection}, we give the analysis for both $\rm ZX$ and $\rm IX$ interactions.

\section{Calibration of $\rm \sqrt{X}$ gate with VZ compensation}

\subsection{Demonstration of the decomposition of a $\frac{\pi}{2}$ gate}

The single-qubit operation that rotates an initial state $\ket{\rm g}$ to the equator of the Bloch sphere can be expressed as:

\begin{equation}
\begin{split}
R &= \left[\begin{array}{cc}
\cos{\frac{\pi}{4}} & -ie^{i\lambda}\sin{\frac{\pi}{4}} \\
-ie^{i\theta}\sin{\frac{\pi}{4}} & e^{i(\lambda+\theta)}\cos{\frac{\pi}{4}}
\end{array}\right].
\end{split}
\end{equation}

A rotation gate around the Z-axis with a rotation angle of $\lambda + \theta$ can be expressed as:

\begin{equation}
\begin{split}
R_z &= \left[\begin{array}{cc}
1 & 0 \\
0 & e^{-i(\lambda+\theta)}
\end{array}\right].
\end{split}
\end{equation}

Applying $R$ and $R_z$ in sequence, we obtain:

\begin{equation}
\begin{split}
R_z R &= \left[\begin{array}{cc}
\cos{\frac{\pi}{4}} & -ie^{i\lambda}\sin{\frac{\pi}{4}} \\
-ie^{-i\lambda}\sin{\frac{\pi}{4}} & \cos{\frac{\pi}{4}}
\end{array}\right]\\
&= \frac{1}{\sqrt 2}\left[\begin{array}{cc}
1 & -ie^{i\lambda} \\
-ie^{-i\lambda} & 1
\end{array}\right]
\end{split}\label{eq1}
\end{equation}

The matrix representation of the operation $e^{-i\frac{\pi}{4}(\cos{\Phi}\sigma_x+\sin{\Phi}\sigma_y)}$ is

\begin{equation}
\begin{split}
R_{\Phi}(\frac{\pi}{2}) &= \frac{1}{\sqrt{2}}\left[\begin{array}{cc}
1 & -ie^{-i\Phi} \\
-ie^{i\Phi} & 1
\end{array}\right].
\end{split}\label{eq2}
\end{equation}

The equations (\ref{eq1}) and (\ref{eq2}) are equivalent as long as we set $\lambda = \Phi$. Thus, the arbitrary single-qubit operation $e^{-i\frac{\pi}{4}(\cos{\Phi}\sigma_x+\sin{\Phi}\sigma_y)}$, which we denote as $R_{\Phi}(\frac{\pi}{2})$, can be decomposed into two operations: $R$ and $R_z$. In this case, $R_{\Phi}(\frac{\pi}{2})$ is simply a $\frac{\pi}{2}$ gate.

\subsection{\label{app: GF} General flow for the calibration of $\rm \sqrt{X}$ gate}

In the experiment, the $\rm \sqrt{X}$ gate is calibrated using the following processes.

1. Prepare the qubit in the initial state $\ket{\text{g}}$, apply the gate to be calibrated, and scan the driving power. The power value corresponding to the final state being $\frac{1}{\sqrt{2}}(\ket{\rm g}+\ket{\rm e})$ is taken as the initial driving amplitude $\Omega_0$.

2. Prepare the qubit in the initial state $\ket{\text{g}}$, and implement two gates to be calibrated, with a VZ compensation in between, having an angle $\phi$. Scan the value of $\phi$, and take the value that corresponds to the maximum probability of the final state being in $\ket{\rm e}$ as the initial angle $\phi_0$ for the VZ compensation.

Take $\Omega_0$ and $\phi_0$ as the initial driving amplitude and VZ compensation. Repeat steps 3 and 4 with increasing values of $n$ until reaching the maximum value of $n$ before decoherence occurs.

3. Prepare the initial state in $\ket{\rm g}$, apply $4n+2$ gates to be calibrated, and scan the driving amplitude. Take the value of the driving amplitude that corresponds to the maximum of $P^{\rm e}$ as the updated $\Omega$.

4. Prepare the initial state in $\ket{\rm g}$, apply 2 gates to be calibrated, followed by 2 conjugate gates of the gate to be calibrated, and repeat this process $n$ times. Scan the value of $\phi$ and take the value that corresponds to the maximum of $P^{\rm g}$ as the updated $\phi$.

\subsection{Calibration of $\rm \sqrt{X}$ gate with dual-DRAG}

We calibrate the $\rm \sqrt{X}$ gate using dual-DRAG to achieve optimal suppression performance. First, we choose typical initial values for the driving amplitude $\Omega$, VZ compensation $\phi$, and the DRAG parameters $\{\alpha, \Delta, -\Delta\}$.

\begin{figure}[htbp]
\includegraphics[scale=1]{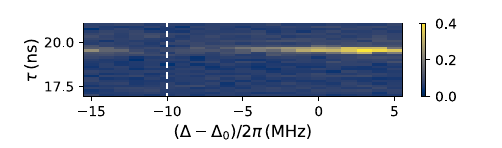}
\caption{\label{suppfig: cali1} Calibration processes with the interference error filter method. Measured $P^{\rm e}$ of $\rm Q_1$ using the same detection circuit as in Fig. 3(b) of the main text, but with dual-DRAG, plotted against the waiting time $\tau$ and the spectral hole-related detuning $\Delta$. The white dashed line indicates the optimized $\Delta$ value used for subsequent experiments.
}
\end{figure}

\begin{figure}[htbp]
\includegraphics[scale=1]{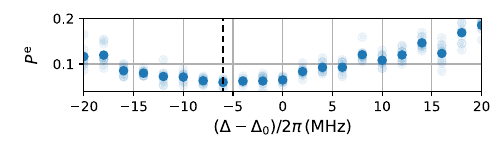}
\caption{\label{suppfig: cali2} Calibration processes with the RB circuit.
Measured $P^{\rm e}$ of the qubit using dual-DRAG with the RB sequence, setting the number of Cliffords to 300, plotted against the spectral hole-related detuning $\Delta$. The black dashed line indicates the optimized $\Delta$ value used for subsequent experiments.
}
\end{figure}

Second, we scan the DRAG point $\Delta$ to find the optimal value using two calibration circuits. The first circuit is the pulse train shown in Fig. 3(b) of the main text. By measuring the spurious excitations around one of the strong peaks, we scan the $\Delta$ value around the predicted value, $\Delta_0$. The minimum value of $P^{\rm e}_1$ corresponds to the optimal $\Delta$, as shown in Fig. \ref{suppfig: cali1}. We also apply the second circuit in another case. In this case, we implement the RB circuit on qubit $\rm Q_0$ with 300 cycles, and measure the population $P^{\rm e}$ of $\rm Q_1$ at the end of the RB sequence (for qubit-TLS system, we will measure the qubit after the qubit reset gate and qubit-TLS iSWAP gate). By scanning the value of $\Delta$, the minimum measured population corresponds to the optimal value of $\Delta$, another case using this calibration method is shown in Fig. \ref{suppfig: cali2}. 

Third, using the optimal DRAG set $\{\alpha, \Delta, -\Delta\}$, we calibrate the $\rm \sqrt{X}$ gate following the calibration flow outlined in Sec. \ref{app: GF}.

\section{Explanations for the detected peaks}\label{sec: app-Detection}

In the experiments for the detection of the spurious transitions, we observed two different groups of peaks with different locations and heights. In this section, we would like to give the detailed analysis for the mechanism of the peaks.

\subsection{ZX crosstalk}\label{app: ZX}
The unitary matrix for the X operation on the left qubit with ZX crosstalk can be written as

\begin{eqnarray}\label{eq: uxzx}
U_{X}^{(ZX)} & \approx & \left[\begin{array}{cccc}
0 & -iA & -i & iB\\
-iA & 0 & -iB & -iC\\
-i & -iB & 0 & iA\\
iB & -iC & iA & 0
\end{array}\right]
\end{eqnarray}
with the approximation of $|g/\Delta_0|\ll1$, where $A =\frac{\Omega^{2}g}{-2\Delta_0(\Delta_0^{2}-\Omega^{2})}+\frac{\Omega^{2}g}{-2\Delta_0(\Delta_0^{2}-\Omega^{2})}e^{-i\Delta_0 T_{g}}$, $B= \frac{\Omega g}{-2(\Delta_0^{2}-\Omega^{2})}+\frac{\Omega g}{-2(\Delta_0^{2}-\Omega^{2})}e^{-i\Delta_0 T_{g}}$, $C=e^{-i\Delta_0 T_{g}}$, $T_g$ is the gate time. Here we assume that, the driving pulse is a square pulse with a real and constant amplitude $\Omega$ and a driving frequency equal to $\Lambda_{\rm eg}$. We dose not distinguish the difference of the frequency detuning between the bare states $\omega_{\rm e}-\omega_{\rm g}$ and dressed states $\Lambda_{\rm ge}-\Lambda_{\rm eg}$. The unitary matrix for the buffer process with the buffer time $T_b$ can be written as

\begin{eqnarray}
U_{bf}\approx\left[\begin{array}{cccc}
1 & 0 & 0 & 0\\
0 & e^{-i\Delta_0 T_{b}} & 0 & 0\\
0 & 0 & 1 & 0\\
0 & 0 & 0 & e^{-i\Delta_0 T_{b}}
\end{array}\right]
\end{eqnarray}
with the approximation of $|g/\Delta_0|\ll1$.

Then we can calculate the unitary matrix elements for the sequence of the operations as with the approximation of $|A|,|B|\ll 1$

\begin{equation}
\label{}
\begin{split}
|[U_{bf}U_{X}^{(ZX)}U_{bf}U_{X}^{(ZX)}]_{[{\rm gg}->{\rm ge}]}^{n(n\geq1)}|
 &\approx |-Be^{-i\Delta_0 T_b}
[1-e^{-i\Delta_0(T_g+T_b)}][1+e^{-i2\Delta_0(T_g+T_b)}+e^{-i4\Delta_0(T_g+T_b)}+...]|\\
 &= |B^{\prime}e^{-i\Delta_0 T_{b}}(1+e^{-i\Delta_0 T_{g}})(1-e^{-i\theta})||\frac{\sin{n\theta}}{\sin{\theta}}|,\\
|[U_{bf}U_{X}^{(ZX)}U_{bf}U_{X}^{(ZX)}]_{[{\rm gg}->{\rm ee}]}^{n(n\geq1)}|
&\approx |-Ae^{-i\Delta_0 T_b}[1-e^{-i\Delta_0 (T_b+T_g)}][1+e^{-i2\Delta_0(T_g+T_b)}+e^{-i4\Delta_0(T_g+T_b)}+...]|\\
 & = |A^{\prime}e^{-i\Delta_0 T_{b}}(1+e^{-i\Delta_0 T_{g}})(1-e^{-i\theta})||\frac{\sin{n\theta}}{\sin{\theta}}|,
\end{split}
\end{equation}
where $A^{\prime}=\frac{\Omega^{2}g}{-2\Delta_0(\Delta_0^{2}-\Omega^{2})}$,
$B^{\prime}=\frac{\Omega g}{-2(\Delta_0^{2}-\Omega^{2})}$,
$\theta=\Delta_0(T_{g}+T_{b})$. The results show that, when $\theta = m \pi$ ($m \in N$), the term of $|\frac{\sin{n\theta}}{\sin{\theta}}|$ is in the peak value. However, when $\theta = 2k \pi$ ($k \in N$), there is $1-e^{-i\theta}=0$, then only when $\theta = (2k+1) \pi$, we can detect the peaks in the states \ket{\rm ge} and \ket{\rm ee}. The buffer time between two adjacent peaks is $\Delta T_b=|2\pi/\Delta_0|$. We can also know that, for our sequence, when the initial state is \ket{\rm gg}, the peaks of \ket{\rm ee} represents the interaction of the first-order process between \ket{\rm gg} and \ket{\rm ge} or \ket{\rm eg} and \ket{\rm ee}, while the peaks of \ket{\rm ge} indicates the second-order interaction between \ket{\rm gg} and \ket{\rm ee} or \ket{\rm ge} and \ket{\rm eg}.

\subsection{IX crosstalk}\label{app: IX}

The unitary matrix for the X operation on the left qubit with IX can be written as

\begin{eqnarray}\label{eq: uxix}
U_{X}^{(IX)}
 & = & \left[\begin{array}{cccc}
0 & 0 & -i & iD\\
0 & 0 & iD & -iC\\
-i & iD & 0 & 0\\
iD & -iC & 0 & 0
\end{array}\right],
\end{eqnarray}
with the approximation of $|g/\Delta_0|\ll1$ and the assumption of a constant driving amplitude, where $D=\nu(1-e^{-i\Delta_0 T_{g}})$,
$C=e^{-i\Delta_0 T_{g}}$. $\nu$ is a constant, the value of which depends on the strength for both of the quantum crosstalk induced IX interaction and the microwave crosstalk. with the approximation of $|D|\ll 1$, the calculated unitary matrix elements for the sequence of the detection circuit are

\begin{equation}
\begin{split}
|[U_{bf}U_{X}^{(IX)}U_{bf}U_{X}^{(IX)}]_{[{\rm gg}->{\rm ge}]}^{n(n\geq1)}|
&\approx|D(1+e^{-i\Delta_0 (T_b+T_g)})[1+e^{-i2\Delta_0 (T_b+T_g)}+e^{-i4\Delta_0 (T_b+T_g)}+...]|\\
 & = |\nu (1-e^{-i\Delta_0 T_{g}})(1+e^{-i\theta})||\frac{\sin{n\theta}}{\sin{\theta}}|\\
 |[U_{bf}U_{X}^{(IX)}U_{bf}U_{X}^{(IX)}]_{[{\rm gg}->{\rm ee}]}^{n(n\geq1)}| 
 & \approx 0
 \end{split}
\end{equation}

Compared with the case of ZX interaction, we can get that, the buffer time between two adjacent peaks is also $\Delta T_b=|2\pi/\Delta_0|$, but only when $\theta = 2k \pi$, the peaks in the states \ket{\rm ge} can be detected. So that, the peaks induced by IX interaction are always resides in the middle of the two peaks induced by the ZX interaction.

\section{Analysis for DRAG corrections}
\label{sec:app-DRAG}

In this section, we would like to give the analysis for the difference between the traditional DRAG with constant detuning and the spectrally balanced DRAG corrections. For the traditional DRAG, the adiabatic transformation is

\begin{eqnarray}
V_c = \exp[-iS_{c}(\frac{g}{\Delta_0}\sigma_y^{\rm gg-ge}+\sigma_y^{\rm gg-eg}+\sigma_y^{\rm ge-ee}-\frac{g}{\Delta_0}\sigma_y^{\rm eg-ee})],
\end{eqnarray}

where $\sigma_y^{j-k}=-i|j\rangle\langle k|+i|k\rangle\langle j|$. Then we can get the results of the first order correction with

\begin{equation}
\begin{split}
S_{c} &= \frac{\Omega^{(0)}}{2\Delta},\\
\Omega^{(1)} &= \Omega^{(0)} - i\frac{\dot{\Omega}^{(0)}}{\Delta}.
\end{split}
\end{equation}

The adiabatic transformation is $H^V=VH_rV^{\dagger}+i\dot{V}V^{\dagger}$, the effective Hamiltonian with the driving pulse $\Omega^{(1)}$ is

\begin{eqnarray*}
H^V\approx\left[\begin{array}{cccc}
 \frac{\eta [\Omega ^{(0)}]^2}{4 \Delta_0 ^2}-\frac{[\Omega ^{(0)}]^2}{2 \Delta_0 } & -\frac{\eta g \Omega ^{(0)}}{2 \Delta_0 ^2}-\frac{g [\Omega ^{(0)}]^3}{4 \Delta_0 ^3} & -\frac{\eta \Omega ^{(0)}}{2 \Delta_0 }-\frac{[\Omega ^{(0)}]^3}{4 \Delta_0 ^2}+\frac{\Omega ^{(0)}}{2} & -\frac{g [\Omega ^{(0)}]^2}{4 \Delta_0 ^2} \\
-\frac{\eta g \Omega ^{(0)}}{2 \Delta_0 ^2}-\frac{g [\Omega ^{(0)}]^3}{4 \Delta_0 ^3} & \frac{\eta [\Omega ^{(0)}]^2 }{4 \Delta_0 ^2}+\eta -\frac{[\Omega ^{(0)}]^2}{2 \Delta_0 }+\Delta_0  & -\frac{\eta g [\Omega ^{(0)}]^2}{2 \Delta_0 ^3}+\frac{3 g [\Omega ^{(0)}]^2}{4 \Delta_0 ^2} & -\frac{\eta \Omega ^{(0)}}{2 \Delta_0 }-\frac{[\Omega ^{(0)}]^3}{4 \Delta_0 ^2}+\frac{\Omega ^{(0)}}{2} \\
 -\frac{\eta \Omega ^{(0)} }{2 \Delta_0 }-\frac{[\Omega ^{(0)}]^3}{4 \Delta_0 ^2}+\frac{\Omega ^{(0)}}{2} & -\frac{\eta g [\Omega ^{(0)}]^2 }{2 \Delta_0 ^3}+\frac{3 g [\Omega ^{(0)}]^2}{4 \Delta_0 ^2} & -\frac{\eta [\Omega ^{(0)}]^2  }{4 \Delta_0 ^2}+\eta +\frac{[\Omega ^{(0)}]^2}{2 \Delta_0 } &\frac{\eta g \Omega ^{(0)} }{2 \Delta_0 ^2}+\frac{g [\Omega ^{(0)}]^3}{4 \Delta_0 ^3} \\
 -\frac{g [\Omega ^{(0)}]^2}{4 \Delta_0 ^2} & -\frac{\delta \Omega ^{(0)} }{2 \Delta_0 }-\frac{[\Omega ^{(0)}]^3}{4 \Delta_0 ^2}+\frac{\Omega ^{(0)}}{2} & \frac{\eta g \Omega ^{(0)}}{2 \Delta_0 ^2}+\frac{g [\Omega ^{(0)}]^3}{4 \Delta_0 ^3} & -\frac{\eta [\Omega ^{(0)}]^2 }{4 \Delta_0 ^2}+2 \eta+\frac{[\Omega ^{(0)}]^2}{2 \Delta_0 }+\Delta_0
\end{array}\right].
\end{eqnarray*}

Because there is $g/\Delta_0\ll1$, we discard the items with the power of $g/\Delta_0$ higher than 1. In our problem, the value of $\Omega ^{(0)}$ is not far less than ${\Delta_0}$, then the term with $\frac{\Omega ^{(0)}}{\Delta_0}$ cannot be discarded simply.  From $H^V[1,1]$ and $H^V[3,3]$, we can get the value of the detuning correction $\eta$ as

\begin{equation}
\eta=-\frac{[\Omega ^{(0)}]^2}{\Delta_0(1- [\Omega ^{(0)}]^2/2 \Delta_0 ^2) }
\end{equation}

According to the experimental and numerical simulation results, the maximum amplitude of the driving pulse is about 30 MHz, then we assume the maximum value of $\Omega^{(0)}/\Delta_0\approx3/4$, the detuning is about $-1.39[\Omega ^{(0)}]^2/\Delta_0$. The item related to the first-order quantum crosstalk is $H^V[1,2]=-0.89[\Omega ^{(0)}]^3 g/2\Delta_0 ^3$ and $H^V[4,3]=-H^V[1,2]$, compared with the original item in $H_r[1,2]$, there is $H^V[1,2]/H_r[1,2]=0.89[\Omega ^{(0)}]^2/\Delta_0 ^2 \approx 0.5$, which is not far less than 1. That is the reason for the loss of efficacy of the traditional DRAG correction. There is one thing to note, even without considering the phase correction and setting $\eta=0$, the value of $H^V[1,2]/H_r[1,2]$ about $0.28$, the suppression for the crosstalk induced transitions is still limited.

For the spectrally balanced DRAG corrections, there is

\begin{eqnarray}
V_b = \exp[-iS_{b1}(\frac{g}{\Delta_0}\sigma_y^{\rm gg-ge}-\frac{g}{\Delta_0}\sigma_y^{\rm eg-ee})-iS_{b2}(\frac{g}{\Delta_0}\sigma_x^{\rm gg-ge}-\frac{g}{\Delta_0}\sigma_x^{\rm eg-ee})],
\end{eqnarray}

where $\sigma_x^{j-k}=|j\rangle\langle k|+|k\rangle\langle j|$. Then we can get the results of the first and second order corrections with

\begin{equation}
\begin{split}
S_{b1} &= \frac{\Omega^{(0)}}{2\Delta},\\
S_{b2} &= -\frac{\dot{\Omega}^{(0)}}{2\Delta^2},\\
\Omega^{(2)} &= \Omega^{(0)} +\frac{\ddot{\Omega}^{(0)}}{\Delta^2}.
\end{split}
\label{app:eq_dual-DRAG}
\end{equation}

After the adiabatic transformation, there are

\begin{eqnarray}
H^V\approx\left[
\begin{array}{cc}
 A & B \\
 C & D
\end{array}
\right],
\end{eqnarray}
where

\begin{eqnarray}
\begin{split}
A &\approx \left[
\begin{array}{cc}
 0 & -\frac{\eta g \Omega^{(0)}}{2 \Delta ^2}+\frac{i \eta g \dot{\Omega}^{(0)}}{2 \Delta ^3}  \\
 -\frac{\eta g \Omega^{(0)}}{2 \Delta ^2}-\frac{i \eta g \dot{\Omega}^{(0)}}{2 \Delta ^3} & \Delta+\eta
\end{array}
\right],\\
B &\approx \left[
\begin{array}{cc}
 \frac{\ddot{\Omega}^{(0)}}{2 \Delta ^2}+\frac{\Omega^{(0)}}{2} & \frac{i g \Omega^{(0)} \dot{\Omega}^{(0)}}{2 \Delta ^3}-\frac{g [\Omega^{(0)}]^2}{2 \Delta ^2}\\
 \frac{i g \Omega^{(0)} \dot{\Omega}^{(0)}}{2 \Delta ^3}+\frac{g [\Omega^{(0)}]^2}{2 \Delta ^2} & \frac{\ddot{\Omega}^{(0)}}{2 \Delta ^2}+\frac{\Omega^{(0)}}{2}
\end{array}
\right],\\
C &\approx \left[
\begin{array}{cc}
\frac{\ddot{\Omega}^{(0)}}{2 \Delta ^2}+\frac{\Omega^{(0)}}{2} & -\frac{i g \Omega^{(0)} \dot{\Omega}^{(0)}}{2 \Delta ^3}+\frac{g [\Omega^{(0)}]^2}{2 \Delta ^2}\\
-\frac{i g \Omega^{(0)} \dot{\Omega}^{(0)}}{2 \Delta ^3}-\frac{g [\Omega^{(0)}]^2}{2 \Delta ^2} & \frac{\ddot{\Omega}^{(0)}}{2 \Delta ^2}+\frac{\Omega^{(0)}}{2}
\end{array}
\right],\\
D &\approx \left[
\begin{array}{cc}
 \eta & \frac{\eta g \Omega^{(0)} }{2 \Delta ^2}-\frac{i \eta g \dot{\Omega}^{(0)}}{2 \Delta ^3} \\
 \frac{\eta g \Omega^{(0)} }{2 \Delta ^2}+\frac{i \eta g \dot{\Omega}^{(0)}}{2 \Delta ^3} & \Delta+2\eta
\end{array}
\right].
\end{split}
\end{eqnarray}

 Here, we also discard the items with the power of $g/\Delta_0$ higher than 1. In this case, from $H^V[1,1]$ and $H^V[3,3]$, we can get the value of detuning with $\eta = 0$. So that, without considering the higher order of $g/\Delta_0$, the items related to the quantum crosstalk are zero.

In reference \cite{hyyppa2024reducing}, based on waveform analysis, the Fourier Ansatz Spectrum Tuning Derivative Removal by Adiabatic Gate (FAST DRAG) and higher-derivative (HD) DRAG methods have been developed to reduce leakage to the second excited state. This is achieved with a gate duration of 6.25 ns and a leakage error of $3 \times 10^{-5}$. In their work, they considered a special case of the HD DRAG method with in-phase and quadrature envelopes, as described by:
\begin{eqnarray}
\Omega_I(t) &=& A[g(t)+\beta_2\ddot{g}(t)]\label{app:eq_inphase}\\
\Omega_Q(t) &=& -\frac{A\beta}{\alpha}[\dot{g}(t)+\beta_2\dddot{g}(t)]\label{app:eq_quadrature}
\end{eqnarray}
The in-phase term in Eq.(\ref{app:eq_inphase}) is similar to our Eq.(\ref{app:eq_dual-DRAG}). The effectiveness of the in-phase envelope in HD DRAG, can be explained by Eq.~(\ref{app:eq_dual-DRAG}). The second derivative term corresponds precisely to the two symmetric DRAG terms, and its coefficient determines the position of the blocking frequency in the spectrum. Therefore, the pulse for $\Omega_I$ is equivalent to an initial pulse of $A g(t)$, plus a pair of spectrally symmetric DRAG terms, one of which resides at the frequency $f_{\rm ef}$. Meanwhile, $\Omega_Q$ represents the third DRAG, which also resides at the frequency $f_{\rm ef}$. This explains why HD DRAG outperforms conventional DRAG in suppressing leakage.

\begin{figure}[htbp]
\includegraphics[scale=1]{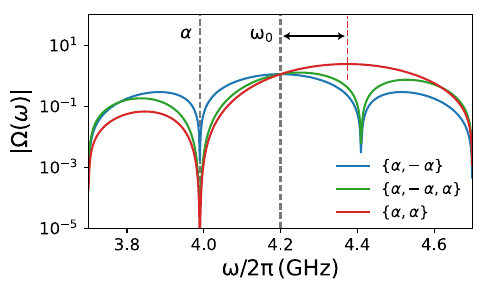}
\caption{\label{suppfig: hddrag} 
Frequency domain representation of the pulse with three different DRAG settings. The central frequency of the initial pulse without DRAG is $\omega_0/2\pi = 4.2~\mathrm{GHz}$, with a pulse duration of $t_g = 6~\mathrm{ns}$ and anharmonicity $\alpha/2\pi = -210~\mathrm{MHz}$. The blue line shows the pulse with dual-DRAG (DRAG set: $\{\alpha, -\alpha\}$), revealing DRAG-blocked frequencies at $\pm \alpha$. The green line corresponds to the pulse with dual-DRAG plus an additional third DRAG at $\alpha$ (DRAG set: $\{\alpha, -\alpha, \alpha\}$). The red line represents the pulse with two DRAGs at the same detuning $\alpha$ (DRAG set: $\{\alpha, \alpha\}$), resulting in a larger frequency shift of approximately 170 MHz, compared to the smaller shift of about 50 MHz seen in the green line.}
\end{figure}

To provide a more detailed explanation of the frequency domain characteristics of different waveforms, we present the frequency domain of a 6-ns pulse with a frequency of 4.2 GHz and an original pulse envelope of $\Omega^{(0)} = \sin^4{\left(\frac{\pi t}{t_g}\right)}$ in Fig.~\ref{suppfig: hddrag}. The blue line represents the dual-DRAG pulse in the frequency domain, with the DRAG detuning occurring at the frequencies of $\pm \alpha$, where $\alpha/2\pi = -210\, \text{MHz}$ is the typical frequency of the anharmonicity. The green line shows the dual-DRAG pulse with the addition of a third DRAG at $\alpha$, which is similar to the third-derivative HD DRAG method discussed in reference \cite{hyyppa2024reducing}. The red line represents a pulse with two DRAGs at the same DRAG detuning of $\alpha$. In this case, the drive frequency is shifted by about 170 MHz, whereas the frequency shift of the pulse shown in green is only about 50 MHz. A smaller frequency shift simplifies the calibration process and improves quantum gate fidelity. The advantage of the pulse with a smaller frequency shift arises from the symmetric DRAG at $-\alpha$, which is the key point discussed in our paper.

In summary, the result from our Eq.~(\ref{app:eq_dual-DRAG}) is derived within the DRAG framework using a full quantum mechanical analysis. This allowed us to obtain both the correction formula and the effective Hamiltonian, which together explain the high performance of HD DRAG. Our findings are consistent with the spectrum optimization approach presented in reference \cite{hyyppa2024reducing}.

\section{Fitting function for the excitation rate}\label{app: fite}

From the equations (\ref{eq: uxzx}) and (\ref{eq: uxix}), we can take the exciting operation on the spectator qubit as a small rotation along the $\sigma_x$ axis with $0<\gamma_e \ll 1$ and 
\begin{equation}
R_x (\gamma_e)=\left[
\begin{array}{cc}
 \sqrt{1-\gamma_e} &  -i\sqrt{\gamma_e}  \\
 -i\sqrt{\gamma_e} &  \sqrt{1-\gamma_e}
\end{array}
\right],
\end{equation}

To extract the excitation rate $\gamma_e$, we assume the qubit goes through a series of channels: $\rho_f = \prod_i O_i (\rho_0)$, where $O_i = C_{\Gamma}\bar{C}_{i}$, $C_{\Gamma}$ is an amplitude damping channel with the damping rate $\Gamma$, and $\bar{C}_{i}= \frac{1}{n}\sum_{j=1}^n  R_z(\alpha_{ij})R_x(\gamma_e)R_{z}(\beta_{ij})$.  $R_{z}$ is the single-qubit rotating gate around the $\sigma_z$ axis with the angle $\alpha$ and $\beta$, which is chosen randomly and when $n \rightarrow +\infty$, there are

\begin{equation}
\begin{split}
\bar{C}_{i} & = \frac{1}{2\pi}\int_{\phi=0}^{2\pi}e^{iH(\phi)\theta}d\phi,\\
H(\phi) & = \cos{\phi}\sigma_x+\sin{\phi}\sigma_y,
\end{split}
\end{equation}
with $\gamma_e=\sin^2{\theta}$. Then there is

\begin{equation}
\begin{split}
\bar{C}_{i}(\rho) &= \cos^2{\theta}\rho+\frac{1}{2}\sin^2{\theta}(\sigma_x\rho\sigma_x + \sigma_y\rho\sigma_y)\\
&=(1-\gamma_e)\rho+\frac{1}{2}\gamma_e(\sigma_x\rho\sigma_x + \sigma_y\rho\sigma_y)
\end{split}
\end{equation}

Initially, there is 

\begin{equation}
\rho_0=\left[
\begin{array}{cc}
 1 & 0 \\
 0 & 0
\end{array}
\right],
\end{equation}

After the first channel $\bar{C}_{1}$, the desity matrix becomes
\begin{equation}
\rho_{0z}=\left[
\begin{array}{cc}
1-\gamma_e & 0 \\
0 & \gamma_e
\end{array}
\right],
\end{equation}

And then after the amplitude damping channel, the density matrix is

\begin{equation}
\rho_{1}=E_1\rho_{0z}E_1^{\dagger}+E_2\rho_{0z}E_2^{\dagger},
\end{equation}
where 
\begin{equation}
E_1=\left[
\begin{array}{cc}
1 & 0 \\
0 & \sqrt{1-\Gamma}
\end{array}
\right],\,
E_2=\left[
\begin{array}{cc}
0 &  \sqrt{\Gamma} \\
0 & 0
\end{array}
\right].
\end{equation}

Then there is
\begin{equation}
\rho_{1}=\left[
\begin{array}{cc}
1-\gamma_e+\gamma_e\Gamma & 0 \\
0 & \gamma_e(1-\Gamma)
\end{array}
\right].
\end{equation}

We assume that, for $\rho_{m-1}$, there is
\begin{equation}
\rho_{m-1}=\left[
\begin{array}{cc}
1-p_1(m-1) & 0 \\
0 & p_1(m-1)
\end{array}
\right],
\end{equation}
where $p_1(m-1)$ is the population on the excited state after $m-1$ Clifford gates.
We can get the population of the excited state in $\rho_{m}$ as

\begin{equation}
\begin{split}
p_1(m) = &r_e-r_e\Gamma +(1-\Gamma-2r_e+2r_e\Gamma)p_1(m-1)\\
=&(r_e-r_e\Gamma)\frac{1-q^m}{1-q}
\end{split}
\end{equation}
for $m=1,2,3...$,  $p_1(0)=0$ and $q=1-\Gamma-2r_e+2r_e\Gamma$.

\section{Explanation of the inter-pulse interference effects}

\begin{figure}[htbp]
\includegraphics[scale=1]{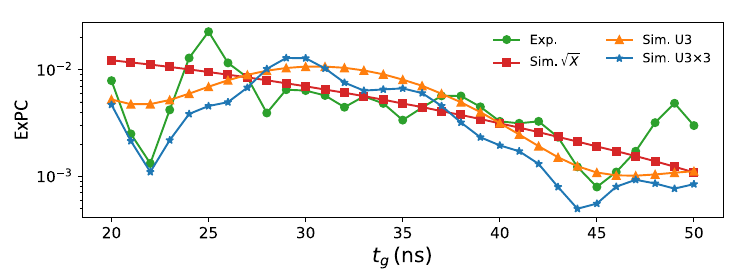}
\caption{\label{suppfig: expc6u} Simulated excitation rates as a function of gate time for single-qubit operations with the DRAG set $\{ \alpha \}$. The green line represents the experimental results. The red line shows twice the excitation rate of the $\rm \sqrt{X}$ gate. The orange line represents the average ExPC from 24 Clifford gates compiled with the U3 scheme. The blue line shows one-third of the excitation rate of averaged single-qubit operations, each consisting of three U3 gates. The average value is computed from 10 random operations.}
\end{figure}

The ExPC as a function of gate time $t_g$ shows an oscillatory behavior in Fig. 4(c) of the main text. This phenomenon can be explained by the inter-pulse interference effect. As shown in Fig. \ref{suppfig: expc6u}, the red line represents twice the simulated excitation rate of the $\rm \sqrt{X}$ gate over gate time, the orange line shows the results from a single U3 gate, and the blue line corresponds to the result from three U3 gates. As the pulse duration increases, the decreasing trend of ExPC for three U3 gates is not monotonic, and oscillations appear, which are attributed to the interference effects between the different pulses. In an RB circuit, which consists of hundreds of U3 gates, the fitted excitation rate is influenced by these inter-pulse interference effects. The practical excitation rate lies within the oscillation range, as shown by the comparison among the simulated ExPCs for $\rm \sqrt{X}$, $\rm U3$, and $\rm U3 \times 3$ in Fig. \ref{suppfig: expc6u}.

\section{Comparison of different DRAG sets}

\begin{figure}[htbp]
\includegraphics[scale=1]{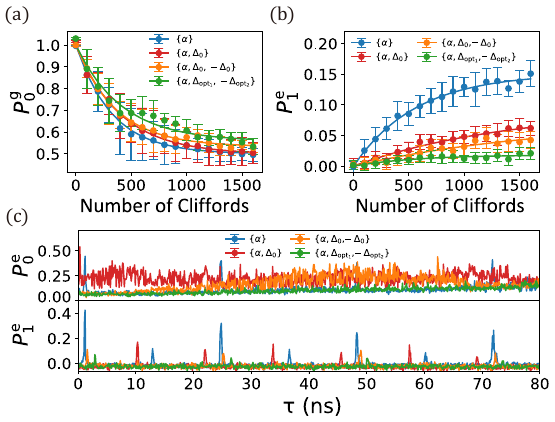}
\caption{\label{suppfig: rbpt}
(a) Randomized benchmarking results for $\rm Q_0$, comparing different DRAG sets.
(b) Simultaneously monitored excited-state populations of $\rm Q_1$. The ExPC values for the DRAG sets $\{\alpha\}$, $\{\alpha, \Delta_0\}$, $\{\alpha, \Delta_0, -\Delta_0\}$, and $\{\alpha, \Delta_{\rm opt}, -\Delta_{\rm opt}\}$ are $(2.7 \pm 0.2) \times 10^{-4}$, $(6.6 \pm 1.0) \times 10^{-5}$, $(4.2 \pm 1.2) \times 10^{-5}$, and $(3.2 \pm 1.2) \times 10^{-5}$, respectively.
(c) Measured excited-state populations of $\rm Q_0$ (top panel) and $\rm Q_1$ (bottom panel), using the detection sequence shown in Fig. 3(b) of the main text, as a function of the waiting time $\tau$. Results are shown for different DRAG sets. All cases include DRAG correction for the leakage transition to the second excited state $\ket{2}$. The number of $\pi$ gate pairs is $N = 50$. Here, $g/2\pi = 0.8~\rm{MHz}$, $\Delta_0/2\pi = 42.5~\rm{MHz}$, and $t_g = 24~\rm{ns}$.}
\end{figure}

To compare the suppression of spurious transitions with different DRAG sets, we present the experimental results for RB and Ramsey interference error filter circuit with four different DRAG sets, as shown in Fig. \ref{suppfig: rbpt}. The ExPC is significantly reduced with the recursive DRAG scheme, and the best suppression occurs with the optimized dual-DRAG, using the DRAG set $\{\alpha, \Delta_{\rm opt}, -\Delta_{\rm opt}\}$, as shown in Fig. \ref{suppfig: rbpt}(b). The significant suppression is further confirmed in Fig. \ref{suppfig: rbpt}(c). The irregular lines in the top panel may be attributed to noise in the system.

\section{Dependence of ExPC over coupling strength}
\begin{figure}[htbp]
\includegraphics[scale=1]{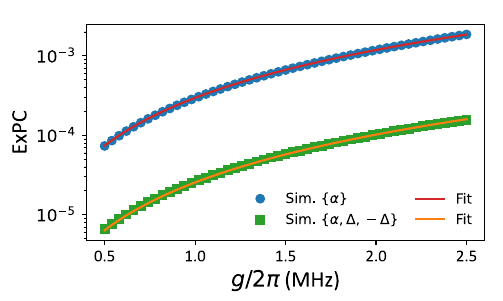}
\caption{\label{suppfig: gsf}
(a) Fitting of the ExPC as a function of coupling strength $g$ using the fitting function $f(x) = ax^2$. The simulated ExPC values are averaged over 24 single-qubit Cliffords implemented with the U3 decomposition. The results with and without dual-DRAG are represented by blue dots and green squares, respectively, with the corresponding fitting results shown as red and orange solid lines.}
\end{figure}

To verify the functional relationship between the ExPC and the coupling strength $g$, we fit the simulated ExPC shown in Fig. 3(g) of the main text using the fitting function $f(x) = ax^2$. The results are shown in Fig. \ref{suppfig: gsf}. The excitation rates are found to be proportional to $g^2$. The fitting parameters are $a_{\rm w/} = 2.6 \times 10^{-5}$ and $a_{\rm w/o} = 3.0 \times 10^{-4}$ with and without dual-DRAG, respectively.

\section{Suppression of the spurious transitions}

\begin{figure}[htbp]
\includegraphics[scale=1]{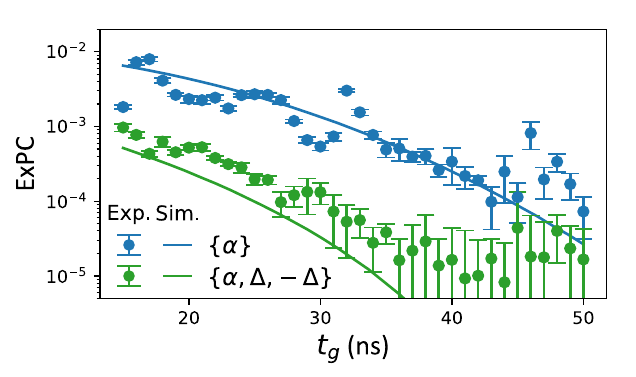}
\caption{\label{suppfig: sctg60} Excitation per Clifford of the TLS for different gate times at $\Delta_0/2\pi = 60~\rm{MHz}$. The solid lines represent twice the simulated $\rm \sqrt{X}$ gate errors.
}
\end{figure}

In Fig. \ref{suppfig: sctg60}, we present the experimental results for TLS excitation rates as a function of gate time ($t_g$) with the detuning $\Delta_0/2\pi = 60~{\rm MHz}$, using the circuit shown in Fig. 4(c) of the main text. Similar to the results presented in the main text, we observe that for gate times in the range $20~\text{ns} \leq t_g \leq 50~\text{ns}$, dual-DRAG reduces TLS excitation rates by an order of magnitude compared to uncorrected pulses. Numerical simulations of single-pulse dynamics successfully reproduce the observed decreasing trend.

\section{Suppression of the spurious transitions with multiple dual-DRAGs}

\begin{figure}[th]
\centering
\includegraphics[scale=1]{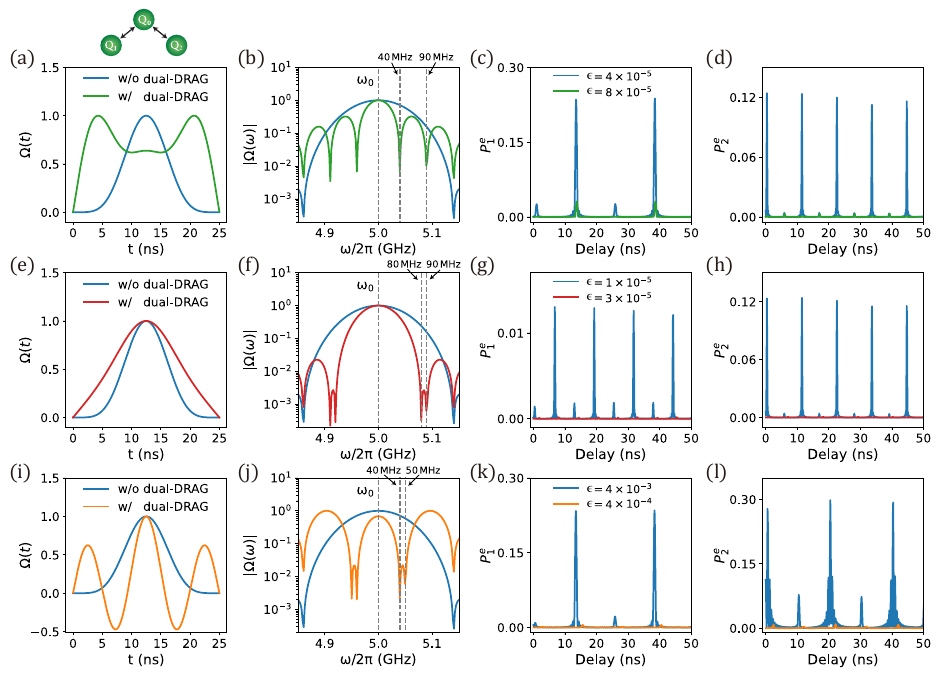}
\caption{\label{fig:two_spectator} The target qubit $\rm Q_0$ couples to two spectator qubits (or TLSs) as shown in the top of the figure, with coupling strengths $g_{01}/2\pi=0.8\,\mathrm{MHz}$ between $\rm Q_0$ and $\rm Q_1$ and $g_{02}/2\pi=2.5\,\mathrm{MHz}$ between $\rm Q_0$ and $\rm Q_2$. The normalized time-domain driving pulses are shown in panels (a, e, i), where blue lines represent pulses without dual-DRAG corrections and green, red, and orange lines show pulses with dual-DRAG corrections, respectively. The corresponding frequency-domain representations are presented in panels (b, f, j). Simulation results for the pulse sequence from Fig. 3(b) (main text) with N=20 are displayed in panels (c, d, g, h, k, l), where the spurious transitions to spectator qubits (or TLSs) are amplified - results for $\rm Q_1$ are shown in (c, g, k) while those for $\rm Q_2$ are in (d, h, l). The detunings are defined as $\Delta_{1} = \omega_1-\omega_0$ and $\Delta_{2} = \omega_2-\omega_0$, with specific values: for panels (a-d), $\Delta_{1}/2\pi=40\,\mathrm{MHz}$ and $\Delta_{2}/2\pi=90\,\mathrm{MHz}$; for (e-h), $\Delta_{1}/2\pi=80\,\mathrm{MHz}$ and $\Delta_{2}/2\pi=90\,\mathrm{MHz}$; and for (i-l), $\Delta_{1}/2\pi=40\,\mathrm{MHz}$ and $\Delta_{2}/2\pi=50\,\mathrm{MHz}$.}
\end{figure}

The dual-DRAG protocol can be generalized to suppress multiple unwanted transitions, though several important considerations must be noted. Multiple DRAG corrections can be applied recursively at different detuning frequencies to eliminate corresponding spectral components of the control pulse. However, standard DRAG pulses may become difficult to calibrate when addressing small detunings due to strong spectral distortions. Our dual-DRAG approach, employing two mirrored DRAG operations for each target frequency, improves this situation and expands the applicable parameter range. This extension comes at the cost of requiring twice as many DRAG operations, with the overall performance strongly dependent on the chosen detuning frequencies.

To illustrate this capability, we examine a system where the target qubit $\rm Q_0$ couples to two spectator qubits $\rm Q_1$ and $\rm Q_2$ with detunings $\Delta_{1} = \omega_{1} - \omega_{0}$ and $\Delta_{2} = \omega_{2} - \omega_{0}$, where $\omega_i$ denotes the frequency of ${\rm Q}_i$ ($i=0,1,2$), as depicted on the top of figure \ref{fig:two_spectator}.

Case 1: Two spectators with detunings $\Delta_{1}/2\pi = 40\,\mathrm{MHz}$ (small) and $\Delta_{2}/2\pi = 90\,\mathrm{MHz}$ (large).
For a gate time $t_g = 25\,\mathrm{ns}$, we apply dual-DRAG corrections at $\pm 40\,\mathrm{MHz}$ and $\pm 90\,\mathrm{MHz}$, followed by fine-tuning. Figure \ref{fig:two_spectator}(a) compares the pulse shapes without and with dual-DRAG in the time domain, while Fig. \ref{fig:two_spectator}(b) shows their frequency-domain counterparts. As demonstrated in Figs. \ref{fig:two_spectator}(c, d), the dual-DRAG corrections suppress spurious transitions to both $\rm Q_1$ and $\rm Q_2$ while maintaining a low gate error with $\epsilon = 4 \times 10^{-5}$.

Case 2: Near-resonant spectators $\Delta_{1}/2\pi = 80\,\mathrm{MHz}$, $\Delta_{2}/2\pi = 90\,\mathrm{MHz}$.
For closely spaced spectators, the dual-DRAG scheme remains effective. The pulse shapes (time and frequency domains) are shown in Figs. \ref{fig:two_spectator}(e) and (f), respectively. Figures \ref{fig:two_spectator}(g, h) confirm significant suppression of spurious transitions to $\rm Q_1$ and $\rm Q_2$, with a gate error of $\epsilon = 1 \times 10^{-5}$.

Case 3: Two small detunings $\Delta_{1}/2\pi = 40\,\mathrm{MHz}$, $\Delta_{2}/2\pi = 50\,\mathrm{MHz}$.
For smaller detunings, spurious transitions are still suppressed, but the driving pulse exhibits stronger frequency-domain distortion, where the amplitude of the side lobes around 4.9\,GHz and 5.1\,GHz exceeds that of the main lobe (Fig. \ref{fig:two_spectator}(j)). While the gate error increases to $4 \times 10^{-3}$, prolonging the gate time can mitigate pulse distortion and restore fidelity.

The performance of multiple dual-DRAG operations is primarily determined by the detuning frequencies involved. While applying multiple dual-DRAG corrections at large detunings (typically $\Delta/2\pi \gtrsim 50$\, MHz for 25-ns pulses) remains straightforward, small-detuning cases ($\Delta/2\pi \lesssim 50$\, MHz) present challenges. Although these can effectively suppress unwanted transitions, they may compromise gate fidelity due to increased spectral distortion. In such cases, gate performance can be improved by optimizing the gate duration.

\section{Leakage suppression with recursive DRAG}\label{app: lkg}

\begin{figure}[htbp]
\centering

\includegraphics[scale=1]{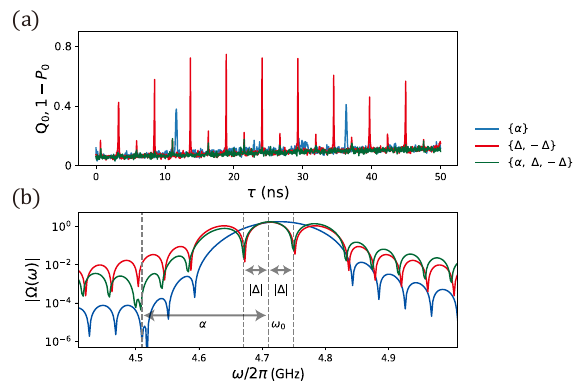}

\caption{\label{fig: discussion} Suppression of more than one transitions with multiple DRAG corrections. (a) Measured sum of the excited states with $\rm Q_0$ when running the detection circuit with three different DRAG sets. (b) Frequency spectrum features of the driving pulses corresponding to different DRAG sets.
}
\end{figure}

Having successfully suppressed spurious transitions using the spectrally balanced DRAG pulse, we now turn our attention to the leakage error on $\rm Q_0$ and talk about the suppression of more than one transitions with multiple DRAG corrections. 
Compared with the conventional DRAG application on the leakage error with the DRAG set of $\{ \alpha \}$, when applying the DRAG correction with the set of $\{ \Delta, -\Delta\}$, the leakage to the excited states of $\rm Q_0$ experiences a sudden increase, as observed by the prominent peaks in FIG. \ref{fig: discussion}(a) with red data line. This increase attributed to the amplification of the driving strength at the frequency of $\omega_{\rm fg}-\omega_{\rm eg}$ due to the DRAG action on $\{ \Delta, -\Delta\}$, as illustrated in FIG. \ref{fig: discussion}(b). To solve this problem, we introduce the third DRAG correction with the DRAG set of $\{ \alpha,\ \Delta, -\Delta\}$, then the leakage associated with the first-order process is significantly suppressed. While there is still some residual leakage error related to the second-order process on $\rm Q_0$ (see minor peaks in FIG. \ref{fig: discussion} with green line), but it is faint enough to be negligible. These results underscore that a driving pulse with multiple DRAG corrections can effectively archieve suppressions on both leakage and spurious transitions.

\end{document}